\documentclass[12pt,a4paper]{article}

\usepackage[T1]{fontenc}
\usepackage[latin1]{inputenc}
\usepackage{color}
\usepackage{graphicx}
\usepackage{amsmath}
\usepackage{amssymb}
\usepackage{subfig}
\usepackage{verbatim}
\usepackage{multirow}
\usepackage{tikz}
\usepackage{hyperref}

\makeatletter \@addtoreset{equation}{section} \makeatother

\makeatletter
\let\old@startsection=\@startsection
\let\oldl@section=\l@section
\renewcommand{\@startsection}[6]{\old@startsection{#1}{#2}{#3}{#4}{#5}{#6\mathversion{bold}}}
\renewcommand{\l@section}[2]{\oldl@section{\mathversion{bold}#1}{#2}}
\makeatother

\makeatletter
\let\old@makecaption=\@makecaption
\def\@makecaption{\small\old@makecaption}
\makeatother

\newcommand{\nn}{\nonumber}

\def\[{\begin{equation}}
\def\]{\end{equation}}

\makeatletter
\def\mr@ignsp#1 {\ifx\:#1\@empty\else #1\expandafter\mr@ignsp\fi}%
\newcommand{\multiref}[1]{\begingroup%\let\protect\string%
\xdef\mr@no@sparg{\expandafter\mr@ignsp#1 \: }%
\def\mr@comma{}%
\@for\mr@refs:=\mr@no@sparg\do{\mr@comma\def\mr@comma{,}\ref{\mr@refs}}%
\endgroup}
\makeatother

\newcommand{\hypref}[2]{\ifx\href\asklfhas #2\else\href{#1}{#2}\fi}
\newcommand{\Secref}[1]{Section~\multiref{#1}}

\newcommand{\Appref}[1]{Appendix~\multiref{#1}}

\newcommand{\Figref}[1]{Figure~\multiref{#1}}

\renewcommand{\eqref}[1]{(\multiref{#1})}

\ifx\href\asklfhas\newcommand{\href}[2]{#2}\fi

\newcommand{\ea}[1]{\begin{align} #1 \end{align}}

\newcommand{\eas}[1]{\begin{align}
\begin{split}#1\end{split}
\end{align}}

\def\Ord{{\cal O}}

\def\a{\alpha}
\def\Eq{\eqref}

\def\e{\epsilon}

\def\op{}%\op

\newcommand*\tcdot{{\mkern -0.5mu\cdot\mkern -0.5mu}}

\newcommand{\sfrac}[2]{{\textstyle\frac{#1}{#2}}}

\newcommand{\atopfrac}[2]{{{#1}\above0pt{#2}}}
\newcommand{\ddelta}{\delta}

\makeatletter
\newlength{\apb@width}
\newcommand{\autoparbox}[2][c]{\settowidth{\apb@width}{#2}\parbox[#1]{\apb@width}{#2}}
\newcommand{\includegraphicsbox}[2][]{\autoparbox{\includegraphics[#1]{#2}}}
\makeatother

\begin{document}
\thispagestyle{empty}

\begin{flushright}\footnotesize
%\texttt{arXiv:xxxx.xxxx}\\
\texttt{HU-EP-18/03}%
\end{flushright}
\vspace{1cm}

\begin{center}%
{\Large\textbf{\mathversion{bold}%
Hidden Conformal Symmetry in Tree-Level Graviton Scattering
}\par}

\vspace{1.2cm}

 \textsc{
Florian Loebbert$^{a}$, 
Matin Mojaza$^{b}$ and
Jan Plefka$^{a}$} \vspace{8mm} \\
\textit{%
$^a$  Institut f\"ur Physik and IRIS Adlershof, Humboldt-Universit\"at zu Berlin,\\ 
Zum Gro{\ss}en Windkanal 6, 12489 Berlin, Germany \\[.2cm]
$^b$ Max-Planck-Institut f\"ur Gravitationsphysik, Albert-Einstein-Institut, \\ 
Am M\"uhlenberg 1, 14476 Potsdam, Germany
} \\

\texttt{\\
\{loebbert,plefka\}@physik.hu-berlin.de\\
mojaza@aei.mpg.de}

%%%%%%%%
\par\vspace{20mm}

\textbf{Abstract} \vspace{5mm}

\begin{minipage}{12.1cm}
We argue that the scattering of gravitons in ordinary Einstein gravity possesses a hidden conformal symmetry at tree level in any number of dimensions. The presence of this conformal symmetry is indicated by the dilaton soft theorem in string theory, and it is reminiscent of the conformal invariance of gluon tree-level amplitudes in four dimensions.
To motivate the underlying prescription, we demonstrate that formulating the conformal symmetry of gluon amplitudes in terms of momenta and polarization vectors requires manifest reversal and cyclic symmetry. 
Similarly, our formulation of the conformal symmetry of graviton amplitudes relies on a manifestly permutation symmetric form of the amplitude function.
\end{minipage}
\end{center}

%%%%%%%%%%%%%%%%%%%%%%%%%%%%%%%%%%%%%%%%%%%%%%%%%%%%%%%%%%%%%%%%%%%%%%%%%%%
%%%%%%%%%%%%%%%%%%%%%%%%%%%%%%%%%%%%%%%%%%%%%%%%%%%%%%%%%%%%%%%%%%%%%%%%%%%
\newpage

\tableofcontents

\bigskip
\noindent\hrulefill
\bigskip

%%%%%%%%%%%%%%%%%%%%%%%%%%%%%%%%%%%%%%%%%%%%%%%%%%%%%%%%%%%%%%%%%%%%%%%%%%%

%%%%%%%%%%%%%%%%%%%%%%%%%%%%%%%%%%%%%%%%%%%%%%%%%%%%%%%%%%%%%%%%%%%%%%%%

\section{Introduction}

Although Yang--Mills (YM) theory and Einstein's theory of gravity both are built upon local symmetries, their
actions look rather different and lead to very distinct quantum properties.
In their perturbative quantization through gluons and  gravitons as 
weak field fluctuations about the vacuum or Minkowski spacetime, respectively, striking similarities in their scattering amplitudes have been discovered. Through a 
unifying UV-completion within string theory,
the KLT relations between open and closed string amplitudes \cite{Kawai:1985xq} allow for a representation of graviton tree-amplitudes as products of color-stripped gluon trees, also at the field theory level. This representation of
gravity as the ``square'' of Yang--Mills theory  was lifted to an entirely new level
through the color-kinematics duality of Bern, Carrasco and Johansson (BCJ) \cite{Bern:2008qj,Bern:2010ue}, which provides a concrete, yet still mysterious, prescription for how  Yang--Mills
amplitudes (at tree and loop-level) may be combined into gravitational amplitudes upon
replacing color degrees of freedom by kinematical ones. 

Clearly symmetries play
a decisive role in constraining the dynamics of quantum field theories and the 
Poincar\'e invariance of scattering amplitudes is a built-in-feature of any 
practical formalism to compute these: Translational symmetry is guaranteed by the
overall momentum conserving delta-function of the amplitude which in turn must be
Lorentz invariant modulo gauge transformations.
Importantly, however, tree-level gluon amplitudes in four dimensions 
are invariant under  the larger group of conformal transformations. While this is a consequence of the classical conformal symmetry of the Yang--Mills action, the conformal symmetry of gluon tree-amplitudes was first fomulated by Witten \cite{Witten:2003nn}, who used an elegant representation
of the conformal generators in spinor-helicity and twistor variables. 

Given the close connection between the conformally invariant gluon tree-amplitudes and
graviton trees, the natural question arises whether the existence of conformal symmetry
for the former leaves any imprint on the structure of the latter. Of course, from an inspection of the Einstein--Hilbert action, a naive conformal symmetry is immediately ruled out due to the dimensionful gravitational coupling in $d>2$. 
However, the existence of \emph{hidden} symmetries in quantum field theories, i.e.~the appearance 
of symmetries at the level of observables which are non-manifest or non-existent 
at the level of the action, has been a recurrent theme in recent years --- just as is the
 case for  the color-kinematics duality discussed above. So it might not be entirely 
misguided to explore the question of a conformal symmetry of graviton tree-amplitudes
beyond $d=2$.

In fact, it has been demonstrated that Einstein gravity in AdS space can be obtained from conformal gravity~\cite{MetsaevDubna, Maldacena:2011mk, Metsaev:2007fq}. 
However, a similar relation does not immediately carry over to flat space.
Further clues towards a hidden symmetry of graviton amplitudes emerged 
from the dicovery of novel subleading soft-graviton theorems \cite{Cachazo:2014fwa}. Extensive works of Strominger et.~al.\ speculate
about the existence of a hidden BMS symmetry \cite{Bondi:1962px,Sachs:1962wk} for \emph{all} massless particle scattering processes
in four dimensions. Along these lines, a recent prescription maps gluon tree-amplitudes in Minkowski space to the celestial sphere at infinity \cite{Pasterski:2017ylz,Schreiber:2017jsr}. The Lorentz symmetry of four-dimensional Minkowski space then acts as the 2d conformal group on the celestial sphere. While the status of this program is still indefinite, soft theorems do indicate that conformal symmetry plays a role for gravity amplitudes. In fact, the
present work was largely motivated by the appearance of the conformal generators of
dilatations and special conformal transformations in the soft theorem for the string theory dilaton
field, as derived in~\cite{Ademollo:1975pf,Shapiro:1975cz,DiVecchia:2015oba,DiVecchia:2016amo,DiVecchia:2016szw,DiVecchia:2017gfi}, and as especially pointed out in~\cite{DiVecchia:2015jaq}.  
The above findings about the soft dilaton are formulated in terms of differential operators
in massless particle momenta and polarizations. Since these variables obey on-shell constraints, e.g.\ $k^2=0$, which generically do not commute with the action of derivatives, we are confronted with an immediate puzzle concerning the interpretation of those results. 
In four dimensions, it would be natural to translate the above statements into spinor-helicity space where the on-shell constraints are unambiguously resolved. However, this route seems not practical here since we require a $d$-dimensional treatment and a scaling dimension different from unity.%
\footnote{Remember that using momentum spinors $\lambda$ and $\bar \lambda$, tree-level Yang--Mills amplitudes in four dimensions are annihilated by the special conformal generator $K^\mu_{\Delta=1}=\frac{\partial^2}{\partial \lambda \partial \bar \lambda}$ \cite{Witten:2003nn} (up to collinear configurations \cite{Bargheer:2009qu}).}

We shall begin our discussion with an analysis of some general features of massless scattering amplitudes and conformal symmetry. Employing a representation of the dilatation and special conformal
generators in terms of differential operators in polarization and momentum vectors (here referred to as \emph{momentum space}) --- and not
in terms of helicity spinors as done in \cite{Witten:2003nn} --- the non-preservation
of the on-shell constraints is demonstrated. This is very subtle, as core properties of
amplitudes such as gauge invariance are only fulfilled on the hypersurface
of the on-shell constraints. We expose the impact of these issues by performing a re-analysis of the conformal symmetry 
of Yang--Mills amplitudes. We then demonstrate that an analogue of the conformal invariance
in four dimensions can nevertheless be found, if an explicit cyclic and reversal symmetrization of the delta-function stripped amplitude is performed. 

We move on to consider graviton scattering and explain how the string theory soft-dilaton limit 
indicates the conformal invariance of graviton amplitudes in ordinary Einstein gravity. A careful analysis suggests that the formulation of this conformal symmetry in momentum space --- in analogy to the Yang--Mills case --- requires a particular representation of the graviton amplitude. 
While the performed analysis is very instructive, at this stage of the paper our statements about the conformal symmetry are still conjectural. 

In the final part, we put our conjecture of the conformal symmetry of graviton amplitudes to the test. We verify that, up to and including multiplicity six, tree-level graviton amplitudes are annihilated by the generators of the conformal algebra. Here, the role of manifest cyclic and reversal symmetry of Yang--Mills amplitudes is taken by full permutation symmetry. That is, in our momentum space formulation, special conformal invariance is found only if we act on the graviton amplitude in a manifestly permutation symmetric form. 
Said differently, the special conformal generator maps the amplitude to the kernel of the full permutation operator. When combined with dilatation and Poincar\'e symmetry, these observations imply the invariance of tree-level graviton amplitudes under the full conformal algebra. The \emph{hidden} character of this symmetry is emphasized by the fact that we require a multiplicity dependent scaling dimension entering the conformal generators.

%%%%%%%%%%%%%%%%%%%%%%%%%%%%%%%%%%%%%%%%%%%%%%%%%%%%%%%%%%%%%%%%%%%%%%%%
%%%%%%%%%%%%%%%%%%%%%%%%%%%%%%%%%%%%%%%%%%%%%%%%%%%%%%%%%%%%%%%%%%%%%%%%

\section{Poincar\'e and Conformal Symmetry in Momentum Space}

In momentum space, amplitudes of $n$ massless particles are described by
a function on the support of an overall momentum conserving $\delta$-function:
\ea{
\mathcal{A}_n (k_1, \ldots, k_n) = \ddelta(P) A_n ( k_1, \ldots, k_n).
\label{eq:strippedamp}
}
Here and throughout this work $\ddelta(P) \equiv \delta^{(d)}(P)$ denotes the $d$-dimensional $\delta$-function with its argument $P$ understood as $P^\mu = \sum_{i=1}^n k_i^\mu$.
The function $A_n$ is the so-called \emph{$\delta$-stripped} or simply \emph{stripped} amplitude.
We will consider massless states  
carrying (symmetric) polarization tensors of the form $\varepsilon^{\mu_1 \cdots \mu_s} = \e^{\mu_1} \cdots \e^{\mu_s}$, and $A_n$ has the property of being linear in $\varepsilon^{\mu_1 \cdots \mu_s}$ for each state, i.e.\ 
\ea{
A_n =  \varepsilon_{i}^{ {\mu_i}_1 \cdots {\mu_i}_s} A_{n, {\mu_i}_1 \cdots {\mu_i}_s}^{(i)}
=  \epsilon_{i}^{ {\mu_i}_1} \cdots \e_{i}^{{\mu_i}_s} A_{n, {\mu_i}_1 \cdots {\mu_i}_s}^{(i)}\, ,
\label{eq:polarizationstripping}
}
where only the linear dependence on the $i$th polarization tensor was exposed. 
The momenta and polarization vectors describing the scattering of massless particles with labels $i=1,\ldots, n$ have to obey the on-mass shell and transversality conditions 
\begin{align}
k_{i}\cdot k_{i} &=0,
&
k_{i}\cdot \epsilon_{i} &= 0\, ,
&\forall i \, .& 
\label{eq:onshellconds}
\end{align}
We will refer to both of these sets of conditions as \emph{on-shell} conditions.
Amplitudes of polarized states described in terms of the $\e_i^\mu$'s are additionally constrainted by gauge invariance; they must be invariant under the gauge transformation $\e_i^\mu \to \e_i^\mu + k_i^\mu$ which is conveniently written as
\ea{
\label{eq:gaugeinvAn}
W_i \mathcal{A}_n = 0 \, ,
}
where we have introduced the generator of gauge transformations
\begin{equation}
\op{W}_i = k_i \cdot \partial_{\epsilon_i}.
\label{defwi}
\end{equation}
The above equation \eqref{eq:gaugeinvAn} is obeyed on the support of the $\delta$-function and on-shell conditions.
We note that $\e_i^\mu$ is not a Lorentz vector (see e.g.\  \cite{Weinberg:1964ew}  and the recent discussions in~\cite{Boels:2009bv, Boels:2016xhc, Arkani-Hamed:2016rak,Roiban:2017iqg}). 

\paragraph{Polarization Tensors of Physical States.}

An elementary physical state should correspond to an irreducible representation of the Lorentz group.
However, for $s$ \emph{even} in \Eq{eq:polarizationstripping} we may also  work with reducible tensor representations, as for instance done in string theory for describing the physical modes of closed strings. For $s=2$, the massless tensor product state characterized by the polarization tensor $\epsilon^{\mu\nu}=\e_i^\mu \e_i^\nu$ describes a multiplet containing both the graviton and a scalar component, the dilaton. The respective $A_{n, \mu \nu}$ represents in this case a reducible stripped amplitude that becomes an amplitude for irreducible representations, once it is contracted with \emph{either} the graviton polarization tensor or the dilaton projector~\cite{Scherk:1974ca,Ademollo:1975pf}. 
Specifically, the symmetric tensor $\varepsilon^{\mu\nu}_i=\e_i^\mu \e_i^\nu$ can be decomposed into a traceless and a trace part  $\varepsilon^{\mu\nu}_i=\varepsilon_\text{graviton}^{\mu\nu}(k_i)+\frac{\e\cdot\e}{\sqrt{d-2}}\,\varepsilon_\text{dilaton}^{\mu\nu}(k_i)$, where 
\ea{\label{eq:epsgravdil}
\varepsilon_{\rm graviton}^{\mu \nu} (k_i)
&= \e_i^\mu \e_i^\nu-
\sfrac{\e\cdot\e}{\sqrt{d-2}} \,\varepsilon_{\rm dilaton}^{\mu \nu} (k_i)\, ,
&
\varepsilon_{\rm dilaton}^{\mu \nu} (k_i)
&= 
\sfrac{1}{\sqrt{d-2}} (\eta^{\mu \nu} - k_i^\mu \bar{k}_i^\nu - k_i^\nu \bar{k}_i^\mu ) \, .
}
Here $\bar k_i$ is an auxiliary momentum which obeys $\bar{k}_i^2 = 0$ and $k_i \cdot \bar{k}_i = 1$ and we have chosen the normalizations such that $(\varepsilon^{\mu \nu}\varepsilon_{\mu \nu})_{\rm dilaton} = 1$ and $\eta_{\mu \nu} \varepsilon_{\rm graviton}^{\mu \nu} = 0$. 
Notice that by imposing $\e_i \cdot \e_i = 0$, we effectively constrain $\varepsilon_i^{\mu \nu} = \varepsilon_{\rm graviton}^{\mu \nu}$.

\paragraph{The Stripped Amplitude.}
To consider $A_n$ separately, i.e.\ to strip off the $\delta$-function, the constraints from Poincar\'e symmetry must be resolved.
This can be achieved by eliminating the momentum of one of the external states using momentum conservation (translational invariance in coordinate space), as well as by imposing the constraints induced via the on-shell conditions of that state (see also~\cite{Boels:2016xhc, Boels:2017gyc} for a detailed discussion). That is, by choosing some state with label $a$ the constraints are resolved by setting
\ea{
k_a^\mu &= - \sum_{i\neq a}^n k_i^\mu \, , &
k_a^2 & =  \sum_{i \neq j \neq a}^n k_i \cdot k_j = 0 \, , &
\e_a \cdot k_a &= - \e_a \cdot \sum_{i\neq a}^n k_i = 0 \, .
\label{MomPoincare}
}
In any expression involving not the full amplitude distribution, but only the stripped amplitude, it will be implicitly assumed that these constraints are enforced on $A_n$ as well as in the end of any manipulation applied to $A_n$. Of course the choice of the state $a$ is arbitrary leading to
an ambiguity in the form of a stripped amplitude.

%%%%%%%%%%

\paragraph{Conformal Transformations in Momentum Space.}
The momentum space generators of the conformal algebra acting on a single leg $i$ of the amplitude read 
\begin{align}
\op{P}_i^{\mu}&= k_{i}^{\mu},&
\op{J}_i^{\mu\nu}&=k_{i}^{\mu}\partial_{k_i}^{\nu}-k_{i}^{\nu}\partial_{k_i}^{\mu}-i S^{\mu\nu}_{i},
\label{ConformalOperators}
\\
\op{D}_{i,\Delta_i} &=  k_i \cdot \partial_{k_i}+\Delta_i,&
\op{K}_{i,\Delta_i}^\mu &=   \sfrac{1}{2} k_i^\mu \partial_{k_i}^2 - (k_i \cdot \partial_{k_i}) \partial_{k_i}^\mu 
- \Delta_i \partial_{k_i}^\mu- i S_i^{\mu \nu} \partial_{k_i, \nu} ,
\nonumber
\end{align}
where for any four-vector $X$
we employ the shorthand notation
\begin{equation}
\partial_{X}^{\mu}:= \frac{\partial}{\partial X_{\mu}}.
\end{equation}
Here $S_i^{\mu \nu}$ is the `spin'  operator 
which, for states with integer spin $s$ whose polarization tensor is symmetric and described by $\varepsilon^{\mu_1 \cdots \mu_s} = \e^{\mu_1} \cdots \e^{\mu_s}$ as in \Eq{eq:polarizationstripping}, can be written as
\begin{equation}
S_i^{\mu \nu} = i \left( \e_i^\mu \partial_{ \e_{i}}^{ \nu} - \e_i^\nu\partial_{ \e_{i}}^{\mu} \right ).
\end{equation}
In the following, we consider $\Delta_i$ to be the same for all legs $i$, thus from hereon $\Delta_i = \Delta$.
The action of the above generators $\op{G}_{i,\Delta}\in\{P_i,J_i,D_{i,\Delta},K_{i,\Delta},S_i\}$ on the whole amplitude is realized via the standard tensor product:
\begin{equation}
\op{G}_\Delta=\sum_{i=1}^n \op{G}_{i,\Delta}.
\end{equation}
The dependence of the generators $\op{D}_\Delta$ and $\op{K}^\mu_\Delta$ on the conformal dimension $\Delta$ is stressed since it will be of particular importance for us. 

%%%%%%%%%%%%%%

\paragraph{Kinematic Hypersurface.}

The above representation of the conformal generators does \emph{not} leave the kinematic
constraint hypersurface of on-shell scattering amplitudes invariant. 
To be explicit, the generator $\op{K}_\Delta^\mu$ does \emph{in general} not  commute with 
$\op{W}_i$, i.e.\
\begin{align}
[K_\Delta^\mu, W_i] &= 
- \partial_{k_i}^\mu W_i + 
(1-\Delta) \partial_{\epsilon_i}^\mu - i S_i^{\mu \nu} \partial_{\epsilon_i, \nu}\, ,
\label{comKW}
\end{align}
nor does it commute with the on-shell conditions $k_i^2 = k_i \cdot \epsilon_i = 0$: 
\begin{align}
[\op{K}_\Delta^\mu, k_i^2] &=  (d- 2 - 2\Delta ) k_i^\mu+ 2 \epsilon_i^\mu W_i,
\\
[\op{K}_\Delta^\mu, k_i\cdot \epsilon_i] &=  \epsilon_i^\mu [ (d- 1 - \Delta )  + \epsilon_i \cdot \partial_{\epsilon_i} ] .
\end{align}
Similarly, overall momentum conservation is generally not  preserved:
\begin{equation}
\label{comOS}
[\op{K}_\Delta^\mu, \ddelta(P)] =  \frac{\partial \ddelta(P)}{\partial P^\nu} \big[ (d - \op{D}_\Delta)\eta^{\mu \nu} + \op{J}^{\mu \nu} \big ].
\end{equation}
Hence, in general $\op{K}_\Delta^\mu$ takes us off the constraint
on-shell surface in momentum  space. For completeness we note that $[K_\Delta^\mu, \epsilon_i\cdot\epsilon_i]=0 $.

The dilatation operator only suffers from not commuting with momentum conservation, i.e.\ we have
\begin{equation}
[\op{D}_\Delta, \ddelta(P)] =  -d \, \ddelta(P).
\label{commDdelta}
\end{equation}
These commutators should, however, be considered on specific amplitudes, 
where (some of) the terms given above may vanish or cancel, as we will see in a moment in the case of $d=4$ YM amplitudes. 

When acting on $\delta$-stripped amplitudes, it is useful to rewrite the conformal generators in terms of differential operators of the Lorentz invariant kinematical variables, $k_i \cdot k_j$, $k_i \cdot \e_j$ and $\e_i \cdot \e_j$, for $i \neq j$, by use of the chain rule, see Appendix~\ref{App:ConformalGenerators} for explicit expressions. We notice that such differential operators were considered recently in \cite{Cheung:2017ems} to describe relations among amplitudes of different theories. 

%%%%%%%%%%%%%%%%%%%%%%%%%%%%%%%%%%%%%%%%%%%%%%%%%%%%%%%%%%%%%%%%%%%%%%%%

\section{Conformal Symmetry of Yang--Mills Amplitudes}
\label{sec:YM}

Yang--Mills theory in four dimensions is classically conformally invariant.
At the level of scattering amplitudes, this was first demonstrated in a manifestly four-dimensional framework in \cite{Witten:2003nn} by using the spinor-helicity formalism with conformal generators represented in spinor space.  Importantly, in this formalism the on-shell conditions 
\eqref{eq:onshellconds}
are explicitly resolved.

Here we wish to understand 
how the conformal invariance of YM scattering amplitudes manifests itself in $d=4$ in the context of a general $d$-dimensional treatment of scattering amplitudes written in momentum space and how it
relates to the on-shell constraints.

%%%%%%%%
\paragraph{Implications of Representation Deficiencies.}

In order to better understand the implications of the on-shell deficiency \Eq{comOS} of the special
conformal generator $\op{K}_\Delta^\mu$, we study the conformal transformations of tree-level YM amplitudes, where we expect to see invariance features when restricting the analysis to four dimensions.
YM amplitudes $\mathcal{A}_n$ can be decomposed into colorless partial amplitudes
\ea{
\mathcal{A}_n (1, \ldots, n) = \ddelta (P) \, g^{n-2} \sum_{{\cal P}(2, \ldots, n)} \text{Tr} [T^{a_1} \cdots T^{a_n}] A_n (1, \ldots, n),
\label{colordecomposition}
}
with $T^{a}$ the color-group generators and $A_n$ denoting from here on a basis of \emph{color-decomposed partial} or simply \emph{partial} amplitudes. The above sum runs over all non-cyclic permutations of the labels $1, \ldots, n$, which can equivalently be expressed as a sum over permutations with one label kept fix.

In four dimensions, the classical scaling dimension of the YM field is  $\Delta=1$.%
\footnote{In $d$ dimensions the canonical scaling dimension of the gluon is 
$\Delta = \frac{d-2}{2}$.}
Independently of the spacetime dimension, the $n$-point stripped amplitude ${A}_n$ is a homogeneous function of the momenta of degree $4-n$.
 Using this, it is easy to show that the dilatation operator annihilates 
the YM amplitude at tree level in $d=4$ and for $\Delta=1$, i.e.
\ea{
\op{D}_{\Delta} \mathcal{A}_n 
=
\big[4-d+n(\Delta-1)\big ]\mathcal{A}_n 
\stackrel{\atopfrac{d = 4}{\Delta=1}}{=}
0,
}
where \Eq{commDdelta} and \Eq{colordecomposition} were used, together with $D_{\Delta=0} A_n = (4-n) A_n$.
We note that dilatation invariance is also obtained
for the multiplicity dependent choice 
$\Delta=\frac{d-4}{n}+1$ in an arbitrary number of dimensions $d$.

Invariance under special conformal transformations is, however, delicate. 
For YM theory, not all of the commutators in \Eq{comKW} and \Eq{comOS} vanish in four dimensions.
Using first 
Lorentz invariance and linearity in the polarization vectors $\epsilon_i^\mu$, as well as on-shell gauge invariance $W_i A_n = 0$, which
are generic $d$-dimensional properties of $A_n$, we find:
\ea{
[K_\Delta^\mu, W_i] A_n 
&=
- \partial_{k_i}^\mu W_i A_n+ 
(1-\Delta) \partial_{\epsilon_i}^\mu A_n
&&\stackrel{{\Delta=1}}{=} - \partial_{k_i}^\mu W_i A_n,
\nn \\
[\op{K}_\Delta^\mu, k_i^2] A_n  
&=  (d- 2 - 2\Delta ) k_i^\mu A_n
&&\stackrel{\atopfrac{d=4}{\Delta=1}}{=} 0,
\nn \\
[\op{K}_\Delta^\mu, k_i\cdot \epsilon_i] A_n 
&=
( d-  \Delta )\epsilon_i^\mu A_n 
&&\stackrel{\atopfrac{d=4}{\Delta=1}}{=}
3\epsilon_i^\mu A_n,
\nn \\
[\op{K}_\Delta^\mu, \ddelta(P)]A_n 
&
= \frac{\partial \ddelta(P)}{\partial P_\mu} \left (d - 4+n(1-\Delta) \right )A_n
&&\stackrel{\atopfrac{d=4}{\Delta=1}}{=}0 \, .
\label{eq:commutators}
}
None of these commutators vanishes generically on the stripped amplitude, and in $d=4$ with $\Delta = 1$ only two out of four commutators vanish. Hence $\op{K}_{\Delta=1}^\mu$ takes the amplitude off the on-shell surface in momentum space, even for $d=4$.
This leads to the question of how conformal invariance in $d=4$ is realized in momentum space.

%%%%%%%%%%%%%

\paragraph{Symmetrization Prescription.}
Poincar\'e invariance implies that the action of $K_\Delta^\mu$ on $A_n$ takes the form%
\footnote{While $\epsilon^\mu$ is not a Lorentz vector, it transforms into $\epsilon^\mu+f(p) p^\mu$ under a Lorentz transformation.}
\ea{
\op{K}_{\Delta}^\mu \, A_n = \sum_{i=1}^n \epsilon_i^\mu F_i + \sum_{i=1}^n k_i^\mu G_i ,
\label{Poincare}
}
where $F_i$ and $G_i$ are some Poincar\'e invariant functions of the kinematic variables, which inherit definite homogeneity degrees in $\epsilon_i^\mu$ and $k_i^\mu$ from the amplitude. 
The coefficients $F_i$ and $G_i$ have the unwanted feature of being dependent on the ambiguity in resolving the Poincar\'e constraints, cf. \Eq{MomPoincare}, presumably as a consequence of \Eq{eq:commutators}. Nevertheless, in the case of explicit lower point examples (to be discussed below) we observe that regardless of this ambiguity
\ea{
\ddelta(P)
 \sum_{i=1}^n k_i^\mu G_i   \stackrel{\Delta=1}{=}0 \, .
\label{Gi}
}
We interpret this as a feature of the underlying conformal symmetry of YM theory in $d=4$, where the ordinary scaling dimension of the gluon becomes $\Delta =\frac{d-2}{2}= 1$. However, we also notice that this relation is valid in any number of dimensions $d$, keeping $\Delta = 1$ fixed.

On the other hand, the explicit examples also reveal  
that the naive application of $\op{K}_{\Delta=1}^\mu$ to $A_n$
does not in general give zero, i.e.\
\ea{
\op{K}_{\Delta=1}^\mu \, A_n = \sum_{i=1}^n \epsilon_i^\mu F_i  \neq 0 \, .
\label{eq:K1neq0}
}
Does this mean that full conformal invariance of tree-level YM amplitudes cannot be seen in momentum space?

At this point, it is useful to note that the stripped partial amplitudes $A_n$ 
inherit cyclic and reversal invariance from the color trace in \Eq{colordecomposition} as follows:
\ea{
\begin{split}
A_n(1,2, \ldots, n) &= A_n (2, \ldots, n, 1), \\
A_n(1,2, \ldots, n) &= (-1)^n A_n (n, \ldots , 2, 1) .
\end{split}
\label{cycref}
}
In order to verify these symmetries, one generically has to resolve momentum conservation and the on-shell conditions, as prescribed by \Eq{MomPoincare}.
However, since $\op{K}_{\Delta}^\mu$ does not commute with the on-shell conditions, we cannot expect these symmetry properties to be preserved by it.
But since $\op{K}_{\Delta}^\mu$ is a fully permutation symmetric differential operator, it seems natural that it should preserve the permutation symmetries of the amplitude.
We can ensure this by manifesting the cyclic and reversal symmetries of $A_n$ by hand, and 
we denote this manipulation by the symbol $\mathcal{C}_n$, i.e.
\ea{
\mathcal{C}_{n}[ A_n ] = \frac{1}{2n}\sum_{\text{Cyc}(1,2, \ldots, n)}\left [ A_n(1,2, \ldots, n) + (-1)^n  A_n(n, \ldots, 2, 1) \right ] \ .
}
%%%%%%%%%%%%%%%
Of course, the stripped amplitude and stripped symmetrized amplitude are identical on the support of momentum conservation and on-shell conditions, i.e.
\ea{
\mathcal{C}_{n}[ A_n ] =  A_n \, .
\label{eq:CnAnequalsAn}
}
In the explicit examples, to be discussed below, we find that
\ea{
\op{K}_{\Delta }^\mu \mathcal{C}_{n}[ A_n ]  = \mathcal{C}_{n}[\op{K}_{\Delta }^\mu A_n ]  \neq \op{K}_{\Delta }^\mu A_n \, ,
}
where the first equality is the trivial statement that $\op{K}_\Delta^{\mu}$, being permutation symmetric, commutes with $\mathcal{C}_n$, while the inequality shows that the naive application of $\op{K}_{\Delta}^\mu$ on $A_n$ does not preserve the permutation properties of $A_n$.
Remarkably, our explicit checks  show that for the specific choice of resolving the Poincar\'e constraints
\ea{
k_n = - \sum_{i=1}^{n-1} k_i \, , \quad k_1 \cdot k_2 = - \sum_{j=3}^{n-1} k_1 \cdot k_j - \sum_{i = 2}^{n-1} \sum_{j>i}^{n-1} k_i\cdot k_j \, , \quad
\e_n\cdot k_1 = - \e_n \cdot \sum_{i=2}^{n-1} k_i \, ,
\label{eq:MomChoice}
}
we systematically find for $n=3,4,5,6$ that
 \ea{
\op{K}_{\Delta=1}^\mu  \, \mathcal{C}_{n}[ A_n] = 0 \, .
 \label{conjecture1}
}
The details for each $n$ are discussed below.
At higher points (i.e.\ for $n>6$) this claim remains conjectural.

The reason why we point out the choice \Eq{eq:MomChoice} is that 
$\op{K}_{\Delta }^\mu \mathcal{C}_{n}[ A_n ] $ still bares some sensitivity to the ambiguity in resolving the Poincar\'e constraints, and \Eq{conjecture1} is not satisfied for all choices.%
\footnote{At least for $n\leq 5$, by summing over all possible ways of resolving the constraints for a fixed $a$ in \Eq{MomPoincare}, $\op{K}_{\Delta }^\mu$ annihilates the expression for any value of $\Delta$.}
Understanding this feature is a nonlinear problem that requires further investigation beyond the scope of this paper, which lies on graviton scattering.
The point we want to make here is that by manifesting the cyclic and reversal symmetries of $A_n$, we find what seems to be the conformal symmetry of YM theory at the level of momentum space amplitudes. It seems plausible that if one could manifest all symmetry properties of $A_n$, such as the photon decoupling property, the Kleiss--Kuijf relations, and perhaps even the BCJ relations, the sensitivity of these statements to the prescription of resolving \Eq{MomPoincare} disappears. In fact, this is what we observe in the graviton case to be discussed later, where all symmetries of the amplitude can be easily implemented as full permutation symmetry.

%%%%
\paragraph{Three-Point Example.}
The three-point stripped YM amplitude takes the form
\ea{
A_3 = (\e_1 \cdot k_2) e_{23}  - (\e_2 \cdot k_1) e_{31} - (\e_3 \cdot k_2) e_{12} \, ,
\label{eq:A3}
}
where we introduced the notation $e_{ij} = \epsilon_i \cdot \epsilon_j$
and resolved the constraints from momentum conservation by imposing
$k_3 = - k_1 - k_2$ and $(\epsilon_3 \cdot k_1) = - (\e_2 \cdot k_2)$.
It is straightforward to compute
\ea{
\op{K}_{\Delta=1}^\mu A_3  = 2 e_{12} \epsilon_3^\mu - 2 e_{23} \epsilon_1^\mu + 2 e_{31} \epsilon_2^\mu \, ,
}
which clearly does not vanish.
Considering instead the cyclic and reversal symmetrized form, which is readily obtained from the above expression,
\ea{
 \mathcal{C}_3 [{A_3}] = \sfrac{1}{2} \e_1 \cdot (k_2 - k_3)\, e_{23} + \sfrac{1}{2}  \e_2 \cdot (k_3- k_1) \,e_{31}  +   \sfrac{1}{2}  \e_3 \cdot (k_1 - k_2) \,e_{12}   \, ,
\label{ua3}
}
it is easily checked that for any value of $\Delta$ we have
\ea{
\op{K}_{\Delta}^\mu \,  \mathcal{C}_3 [{A_3}]  = 0 .
\label{Kua3}
}
The three-point amplitude is of course very special (e.g.\ it vanishes for real momenta).
However, as we will see next, the four-, five-, and six-point amplitudes expose this symmetry through the same procedure, but where $\Delta = 1$ becomes a crucial choice.
%%%%
\paragraph{Four-Point Example.}
The expression for $\mathcal{A}_4$ will not be provided here, but can be straightforwardly computed from just four Feynman diagrams using textbook prescriptions. The partial stripped amplitude $A_4$ is obtained by imposing \Eq{eq:MomChoice}, specifically
$k_4 = - k_1 - k_2 - k_3$, $s_{12} = - s_{13} - s_{23}$ and $\epsilon_4 \cdot k_1 = - \epsilon_4 \cdot (k_2 + k_3)$.
We then find
\ea{
\op{K}_{\Delta=1}^\mu \,  {A_4} = \sum_{i=1}^4 \e_i^\mu F_i \neq 0,
}
where $F_i$ are some nonzero functions of the kinematic variables. For $\Delta\neq 1$, also terms proportional to $k_i^\mu$ contribute. 
By considering instead the cyclic and reversal symmetrized form we find as conjectured
\ea{
\op{K}_{\Delta =1}^\mu \,  \mathcal{C}_4 [{A_4}]   = 0 \, , 
}
which holds only for $\Delta=1$. It moreover turns out that the coefficients of the $k_i^\mu$ here vanish, not only for $\Delta =1$, but for any value of $\Delta$. This is, as in the three-point example, a special property at four points that we do not find at higher points.

Curiously, this invariance is also seen when writing $A_4$ in a manifestly gauge invariant form.
This form is obtained by employing the so-called $t_8$-tensor~\cite{Schwarz:1982jn} and reads
\ea{
A_4 = \frac{4}{s_{12}s_{23}} t_{8, \mu_1 \nu_1, \ldots, \mu_4 \nu_4} k_1^{\mu_1} \cdots k_4^{\mu_4} \epsilon_1^{\nu_1} \cdots \epsilon_4^{\nu_4},
\label{t8expression}
}
with $s_{ij} = 2 k_i \cdot k_j$ and $t_8$ being a tensor with symmetry under exchange of any pairs of indices $\{\mu_i \nu_i\}$ and antisymmetry under the exchange $\mu_i \leftrightarrow \nu_i$ in each pair, making it manifestly gauge invariant. Here we emphasize that $k_4 = - (k_1 + k_2 + k_3)$.
In addition, the numerator has manifest cyclic and reversal symmetry (in fact, it has full permutation symmetry), but the denominator does not manifest these properties. Nevertheless, we find that in this form $A_4$ is annihilated by $\op{K}_{\Delta}^\mu$ iff $\Delta=1$. 
An additional, possibly related curiosity is that the action of $\op{K}_{\Delta}^\mu$ on $A_4$ as given above is gauge invariant iff $\Delta=1$. This can be shown using only the symmetry properties of $t_8$, i.e.\ without using on-shell conditions. The absence of a similarly compact expression for higher point amplitudes suggests these observations to reflect a special symmetry at four points.

%%%%
\paragraph{Five-Point Example.}
Depending on parametrization, the five-point YM amplitude consists of some 400 individual terms.%
\footnote{We thank C. Mafra and O. Schlotterer, as well as J. Bourjaily for providing us with 
explicit momentum space expressions for YM tree-amplitudes. The lower $n$-point espressions have been fully exposed in \cite{Mafra:2010jq,Bjerrum-Bohr:2016juj}.}
Imposing \Eq{eq:MomChoice}, we again find that while \Eq{Gi} is satisfied, we still have
\ea{
\op{K}_{\Delta=1}^\mu \,  A_5 =\sum_{i=1}^5 \e_i^\mu F_i \neq 0 \, .
}
By performing the cyclic and reversal symmetrization as in the previous examples, 
we then find as conjectured 
\ea{
\op{K}_{\Delta=1}^\mu \, \mathcal{C}_5 [ {A_5}]  = 0 \, .
}
This latter check has been performed numerically only and is satisfied iff $\Delta=1$.

%%%%
\paragraph{Six-Point Example.}
Up to five points, the Yang--Mills tree amplitude in four dimensions is either MHV (maximally helicity violating) or $\overline{\text{MHV}}$. To ensure that the manifestation of conformal symmetry, as found in the previous examples, is not just a special feature of this helicity configuration, we have also explicitly checked the six-point case. Depending on parametrization it consists of some 6-7000 individual terms. Also in this case, while we do find \Eq{Gi} to be satisfied, $A_6$ defined by imposing \Eq{eq:MomChoice} is again generically not   annihilated by $\op{K}_{\Delta=1}^\mu$. After manifesting the cyclic and reversal symmetries as in the previous examples, we nevertheless find our conjecture \Eq{conjecture1} to hold iff $\Delta=1$. The checks here have all been performed numerically.

%%%%%%%%%%%%%%%%%%%%%%%%%%%%%%%%%%%%%%%%%%%%%%%%%%%%%%%%%%%%%
\section{Soft Dilatons, Conformal Generators and Graviton Scattering}
\label{sec:softtheorem}

A hint on special conformal invariance of tree-level graviton amplitudes in field theory is observed when considering the soft behavior of massless closed strings in any string theory (bosonic, heterotic, or in the NS-NS sector of superstrings). These limits have recently been studied in \cite{DiVecchia:2015oba,DiVecchia:2016amo,DiVecchia:2016szw,DiVecchia:2017gfi}. The results of those works, important to us, are the following.
 Consider for simplicity scattering amplitudes in bosonic string theory using the operator formalism (the same analysis and result applies in the RNS formalism of superstrings).
The $n$-point bosonic string amplitude of massless closed states carrying momenta $k_i$ and polarizations $\e_i, \bar{\e}_i$ can be written as
\begin{equation}
\mathcal{ M}_n^{\alpha'}= \ddelta(P) \widetilde M_n^{\a'},
\end{equation}
with
\eas{
\widetilde M_n^{\a'} = & \frac{8\pi}{\alpha'}\left (\frac{\kappa_d}{2\pi}\right )^{n-2} 
\int \frac{\prod_{i=1}^n d^2z_i }{dV_{abc}}   \prod_{i<j} |z_i - z_j |^{\alpha' k_i k_j}   \\
& \times \int \prod_{i=1}^n d\theta_i  \exp \left[\sum_{i<j}  \frac{(\theta_i \epsilon_i) \cdot (\theta_j\epsilon_j)}{(z_i - z_j)^2}   + \sqrt{\frac{\alpha'}{2}} \sum_{i \neq j} \frac{ (\theta_i \epsilon_i)\cdot k_j }{z_i - z_j}  \right]  \\
& \times
\int \prod_{i=1}^{n} d {\bar{\theta}}_i 
 \exp \left[\sum_{i<j}  \frac{({\bar{\theta}}_i {\bar{\epsilon}}_i ) \cdot ({\bar{\theta}}_j{\bar{\epsilon}}_j)}{({\bar{z}}_i - {\bar{z}}_j)^2} + \sqrt{\frac{\alpha'}{2}} \sum_{i \neq j} \frac{ ({\bar{\theta}}_i {\bar{\epsilon}}_i )\cdot k_j }{{\bar{z}}_i - {\bar{z}}_j}  \right] ,
\label{Mn}
}
where $P= \sum_{i=1}^n k_i$, $\kappa_d$ is the $d$-dimensional gravitational constant, and integration is over the Koba--Nielsen variables $z_i$, modulo $\mathrm{SL}(2, \mathbb{C})$ symmetry which is fixed by
\ea{
dV_{abc} = \frac{d^2 z_a d^2 z_b d^2 z_c }{|z_a-z_b|^2 |z_b-z_c|^2 |z_c-z_d|^2} \, .
}
The points $z_a, z_b, z_c$ can be fixed to any point in the complex plane
and the indices $a,b,c$ are any three from the set $\{ 1, \ldots, n\}$.
The $\theta_i$, $\bar{\theta}_i$ are Gra{\ss}mann variables introduced to exponentiate the integrand (the $\epsilon_i$ are thus also Gra{\ss}mann). 

By explicitly calculating the integral over one of the $z_i$ in the limit where the corresponding momentum is soft compared to the other momenta, 
it has been shown~\cite{DiVecchia:2015oba,DiVecchia:2016amo,DiVecchia:2016szw}
that the soft behavior of a symmetrically polarized, massless closed string state 
can be described by the expression
\ea{
 \mathcal{M}_{n+1}^{\alpha'} (k_1, \ldots, k_n, q)
 =&
 \ddelta(P + q)\left ( \op{S} \, \widetilde M_n^{{\a'}}(k_1, \ldots, k_n)  + \Ord(q^2)\right),
\label{softtheorem}
}
where
\ea{\label{eq:opS}
\op{S} &=\kappa_d \, \varepsilon_{q, \mu \nu}
 \sum_{i=1}^n  
 \left [ \frac{k_i^\mu k_i^\nu}{k_i \cdot q} + \frac{ k_i^{\mu} q_\rho (- i J_i^{\nu \rho})}{k_i \cdot q}
-
\frac{q_\rho q_\sigma  }{2 k_i \cdot q}  : J_i^{\mu \rho}  J_i^{\nu \sigma} :
+\,  \Ord(\alpha' q) \right ] \, , 
\\[2mm]
J_i^{\mu \nu} &=
 i \left (k_i^\mu \partial_{k_{i}}^{ \nu} - k_i^\nu \partial_{k_{i}}^{{ \mu}} \right )
+ i \left (\e_i^\mu \partial_{\e_{i}}^{{ \nu}} - \e_i^\nu \partial_{\e_{i}}^{{ \mu}} \right )
+ i \left (\bar{\e}_i^\mu \partial_{\bar{\e}_{i}}^{{ \nu}} - \bar{\e}_i^\nu \partial_{\bar{\e}_{i}}^{{ \mu}} \right ) \, .
}
The soft, symmetrically polarized state carries momentum $q$ with the polarization tensor $\varepsilon_q^{\mu \nu} = \e_q^{(\mu} \bar{\e}_q^{\nu)}$. Normal ordering $: \ :$ means that all derivatives act to the right.
In bosonic string theory, but not in superstring theory, an additional operator contributes to $\op{S}$ at order $q$ (cf.\ \cite{DiVecchia:2015oba}), which is proportional to the inverse string tension $\alpha'$ and which is not of interest here.
The operators act on $\widetilde M_n^{{\a'}}$, which is the integral representation in \Eq{Mn}, but
\emph{before} imposing momentum conservation, hence the tilde (see Appendix~\ref{threepointresolution} for further discussion). 
This remark is important, because both sides of \Eq{softtheorem} involve the same $(n+1)$-point $\delta$-function. Eq.~\Eq{softtheorem} is therefore not a soft theorem in the usual sense, where amplitudes map to amplitudes and this will be important to us.
This `soft theorem' is instead stating that the integration over the moduli of the soft state in $ M_{n+1}^{{\a'}}$ can, for the first three orders in the $q$-expansion, be written as a differential equation on $\widetilde M_n^{{\a'}}$ with the overall $(n+1)$-point momentum conservation from $M_{n+1}^{{\a'}}$ being kept outside. 
We would like to understand to what extent \Eq{softtheorem} can be interpreted in terms of amplitudes, in particular in the field theory limit where $\a' \to 0$.

It is useful to note that the theorem in \Eq{softtheorem} was also derived in field theory from
on-shell gauge invariance of $\mathcal{M}_{n+1}$, 
under the assumption that no terms proportional to $\eta_{\mu \nu}\varepsilon_q^{\mu \nu}$ appear%
\footnote{
This assumption holds at tree-level to all orders, in fact, as a corollary of the 
KLT relations~\cite{Kawai:1985xq}, which state that the amplitude factorizes into two copies of Yang--Mills amplitudes.
 }
 up to $\Ord(q^2)$,
where instead of $\widetilde{M}_n^{\a'}$ an unknown, but in principle calculable, $n$-point \emph{current} with one leg off shell enters~\cite{DiVecchia:2015jaq}.
Consistency of the two different methods of obtaining \Eq{softtheorem} thus indicates 
that there is a well defined way of taking the $\a' \to 0$ limit of $\widetilde M_n^{\a'}$
entering \Eq{softtheorem}, but neither approach immediately yields relations among amplitudes.
For obtaining these, there are at least two ways to proceed: One is to analyze carefully $\widetilde M_n^{\a'}$, as it enters in \Eq{softtheorem}, and its field theory limit. We provide in Appendix~\ref{threepointresolution} such an analysis in the simplest case of $\widetilde M_3^{\a'}$, which clarifies the problems and shows how the above equations can be turned into a statement for a representation of the three-point stripped amplitude $M_3^{\a'}$.
Another way is to commute the $\delta$-function through the soft operators to obtain a statement on the level of amplitudes in analogy to the discussion of
\cite{Broedel:2014fsa}. This is the path we will take here. Crucial for the success of this approach is that after expanding $\ddelta(P+q)$ as well as the prefactor $\widetilde M_{n+1}$ in powers of $q$, and after pulling the soft operator through the resulting $\delta$-function $\ddelta(P)$, all derivatives of the $\delta$-function cancel so that one generically ends up with an expression including the $n$-point amplitude:
\ea{
\ddelta(P + q)  M_{n+1}^{{\a'}}(k_1, \ldots, k_n,q) 
=& \Big[\ddelta(P) + \big(q\cdot \partial_P \ddelta(P)\big) \Big]\op{S}\,\widetilde M_n^{\alpha'}(k_1, \ldots, k_n)+\mathcal{O}(q^2)
\nonumber\\
=&\op{\widetilde S} \,\ddelta(P) M_n^{\alpha'}(k_1, \ldots, k_n)+\mathcal{O}(q^2).
\label{softgraviton}
}
Because of  $n$-point momentum conservation, $\widetilde M_n^{{\a'}}$ becomes the $\delta$-stripped scattering amplitude $M_n^{\alpha'}$ in the last line. Note that the nontriviality in the above expression lies in the existence of a differential operator $\op{\widetilde S}$ which satisfies \eqref{softgraviton}.

%%%%%%%%%%%%%%%%%%%%%%%

\paragraph{Soft Graviton Theorem.}

The field theory limit of \Eq{softtheorem} consistently reproduces the tree-level soft theorem of the graviton~\cite{Weinberg:1964ew,Gross:1968in}, including the recently discovered subsubleading term~\cite{Cachazo:2014fwa,Broedel:2014fsa,Bern:2014vva} for any number of dimensions $d$.
This can be easily seen, by first noting that for the graviton, the normal ordering symbol in the operator $S$ in (\ref{eq:opS}) can be removed due to the tensor properties of $\varepsilon_{\rm graviton}^{\mu \nu}$, yielding immediately the known form for the graviton soft operator, and secondly by noting that this operator has been shown to have the property~\cite{Broedel:2014fsa}
\ea{
\ddelta(P+q) S_{\rm graviton} = S_{\rm graviton} \ddelta(P) + \Ord(q^2) \, .
}
Hence, the problem of understanding (\ref{softtheorem}) in terms of amplitudes is immediately solved in the case of the graviton and, in this case, (\ref{softgraviton}) holds for $\op{\widetilde S} = S$.

%%%%%%%%%%%%%%%%%%%%%%%

\paragraph{Soft Dilaton Theorem.}

\begin{figure}
\begin{center}
\includegraphicsbox{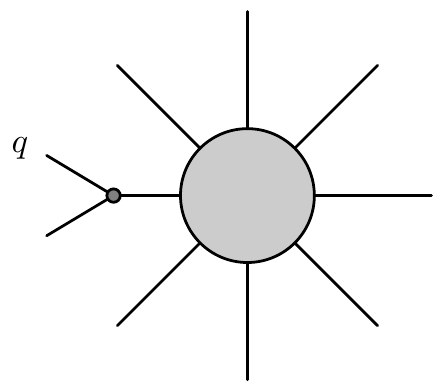}
\caption{Emission of a soft particle with momentum $q$ from a hard external leg.
}
\label{fig:3point}
\end{center}
\end{figure}

The soft behavior of the dilaton is obtained by replacing the polarization tensor of the soft state with the projector (cf.\ \cite{Scherk:1974ca,Ademollo:1975pf} and the discussion around \eqref{eq:epsgravdil})
\ea{
\varepsilon_{\rm dilaton}^{\mu \nu} = \frac{1}{\sqrt{d-2}} (\eta^{\mu \nu} - q^\mu \bar{q}^\nu - q^\nu \bar{q}^\mu ) \, , \quad \text{for} \ \bar{q}^2 = 0 \ \text{and}\  q \cdot \bar{q} = 1.
\label{dilatonprojector}
}
Assuming for simplicity that all hard states are either gravitons or dilatons, such that we can take $\bar{\e}_i \to \e_i$, the `soft theorem' in \Eq{softtheorem} takes the form~\cite{DiVecchia:2015oba,DiVecchia:2016amo,DiVecchia:2016szw}: 
\begin{align}
\mathcal{ M}_{n+\phi}^{\alpha'}(k_1,.., k_n, q) 
=
 \sfrac{\kappa_d}{\sqrt{d-2}}\Big[\ddelta(P) 
S_\delta
+ 
\big(q\cdot \partial_P \ddelta(P)\big)S_{\delta'}
+(S_W+\op{S}_V)
\ddelta(P)  \Big ] \widetilde M_n^{{\a'}} 
+{\cal{O}} (q^2 ) ,
\label{DilatonSoftTheorem}
\end{align}
where the coefficients of $\delta(P)$ and $\delta'(P)$ are given by the local operators
\begin{align}
S_\delta
&=2 - \op{D}_{\Delta=0} +q_\mu \op{K}_{\Delta=0}^\mu ,
&
S_{\delta'}
&=
2 - \op{D}_{\Delta=0},
\label{eq:SdSdp}
\end{align}
as well as the non-local operators
\begin{align}
S_W
&= -\sum_{i=1}^n\frac{q \cdot \e_i}{k_i \cdot q}  
 \left (1 + q\cdot \partial_{k_i} \right ) W_i ,
 &
\op{S}_V
&= 
\sum_{i=1}^{n} \frac{q_\rho q^\sigma}{ 2 k_i\cdot q} \left(      
 S_i^{ \rho \mu  }  S_{i, {\mu \sigma } }
  +   d \,  \e_{i}^{\rho} 
\partial_{\e_{i },\sigma} 
\right)\, .
 \label{eq:S12}
\end{align}
Here $W_{i}$ is the generator of gauge transformations \Eq{defwi}.
To obtain this result, momentum conservation was used\footnote{
The factor $2$ in the soft operator $S_{\delta}$ arises by use of momentum conservation from the leading $q^{-1}$ term in \Eq{softtheorem}; i.e. the projector \Eq{dilatonprojector} on the leading term gives 
$-2 \ddelta(P+q) P \cdot \bar{q} = 2 \ddelta(P+q) q\cdot \bar{q} = 2 \ddelta(P+q) $.
This can also be obtained by first expanding the $\delta$-function 
$-2 \ddelta(P+q) P \cdot \bar{q} = -2 \ddelta(P)  P \cdot \bar{q}
- 2 (q\cdot \partial_P \ddelta(P)  ) P \cdot \bar{q}$. The first term is zero because $P=0$, while for the second term we can use the identity $P^\mu \partial_P^\nu \ddelta(P) = - \eta^{\mu \nu} \ddelta(P)$, leading to the same result.}
 as well as the on-shell conditions \mbox{$q^2=0$}, \mbox{$k_i^2= k_i \cdot \e_i = 0$} and Lorentz invariance of $\widetilde M_n^{{\a'}} $.
The operator $S_W$ was moved to the left of the $\delta$-function by making use of the following, easily checked relation:
 \ea{
\ddelta (P+q) S_W =  S_W \ddelta (P)  + \Ord(q^2) .
 } 
Notice that $S_W$ contains the operator $\partial_{k_i}W_i$
which annihilates \emph{manifestly} gauge invariant expressions.
For the contributions to \eqref{DilatonSoftTheorem} where $\widetilde M_n^{\a'}$ is directly multiplied by $\ddelta(P)$, the integral expression defined in \eqref{Mn} becomes the proper $\delta$-stripped amplitude, i.e.\
\ea{ 
(S_W+\op{S}_V)
\ddelta(P) \widetilde M_n^{{\a'}}  = (S_W + \op{S}_V) \mathcal{M}_n^{\a'} \, .
}
From an analysis of Feynmann diagrams, these terms
must correspond to diagrams of the type in \Figref{fig:3point},
due to the propagator-pole at $k_i\cdot q=0$, as will be further discussed below. The splitting into $S_W$ and $\op{S}_V$ is done, because as we will argue in a moment, we can effectively set
\begin{align}
&S_W \mathcal{M}_n^{\a'}=0+{\cal{O}} (q^2 ) .
\end{align}

Remarkably, the operators $\op{D}_{\Delta=0}$ and $ \op{K}_{\Delta=0}^\mu$ appearing in \eqref{eq:SdSdp} after the projection onto the dilaton, are exactly the conformal generators defined in \Eq{ConformalOperators} for $\Delta=0$, as first pointed out in~\cite{DiVecchia:2015jaq}.
This gives a first glimpse at the role played by the conformal algebra in the context of graviton scattering, on which we will elaborate below.

The leading part of the above soft theorem was already understood in works dating back to the 1970s in relation to scale renormalization in string theory~\cite{Ademollo:1975pf,Shapiro:1975cz}.
The full soft behavior (\ref{DilatonSoftTheorem}) was derived in~\cite{DiVecchia:2015oba,DiVecchia:2016amo,DiVecchia:2016szw} for different string theory setups and in~\cite{DiVecchia:2015jaq} in field theory. 
As shown in~\cite{DiVecchia:2016amo}, the expression (\ref{DilatonSoftTheorem}) is universal; i.e.\ it describes the soft behavior of the dilaton in any string theory, meaning that the operators of order $\alpha' q $ in (\ref{eq:opS}) vanish for the dilaton.
There, however, the  operator $S_W$ was immediately dropped, but we wish here to scrutinize the arguments for this.

%%%%%%

\paragraph{The Operator $S_W$.}

If one assumes that the amplitude $\mathcal{M}_n^{{\a'}}$ takes a \emph{manifestly} gauge invariant form, on which the action of $W_i=k_i\cdot\partial_{\epsilon_i}$ is identically zero (i.e.\ without using on-shell constraints), the operator $S_W$ can be immediately dropped. 
However, such a representation of the amplitude is generically not known, 
and the argument seems not particularly useful for us.
Let us therefore give an alternative argument for why this term should drop out in \eqref{DilatonSoftTheorem}. Notice that we can write
\begin{equation}
S_W
\ddelta(P) \widetilde M_n^{{\a'}} 
+{\cal{O}} (q^2 )
=S_W
\mathcal{M}_n^{\a'}
+{\cal{O}} (q^2 )
=
-\sum_{i=1}^n\frac{q \cdot \e_i}{k_i \cdot q}  
e^{q\cdot \partial_{k_i}}
k_i \cdot \partial_{ \e_i}
\mathcal{M}_n^{\a'}
 +{\cal{O}} (q^2 ),
\label{eq:eq312}
\end{equation}
where we used that $\ddelta(P) \widetilde M_n^{{\a'}}  = \mathcal{M}_n^{\a'}$, as well as the expansion of the momentum shift operator:  
$e^{q\cdot \partial_{k_i}}=1 + q\cdot \partial_{k_i}+\mathcal{O}(q^2)$.
Evaluating the shift operator, we can thus write \eqref{eq:eq312} in the form
\ea{
-\sum_{i=1}^n\frac{q \cdot \e_i}{k_i \cdot q}  
(k_i+q) \cdot \partial_{ \e_i}  
\mathcal{M}_n^{\a'}(k_1, \ldots, k_i + q, \ldots , k_n)
+ \Ord(q^2)
 = 0+ \Ord(q^2).
 \label{resummation}
 } 
The final zero follows from on-shell gauge invariance of the amplitude $\mathcal{M}_n^{\a'}$, which does not necessitate manifest gauge invariance.
Notice, however, that the rewriting in \eqref{resummation} relies on the addition of an infinite number of terms of higher order in $q$ which complete the original expression to the full gauge invariance condition. 
The potential danger of such a resummation is that the order-by-order gauge invariance in the $q$-expansion is spoiled. This is further discussed in Appendix~\ref{App:gaugeinvariance}.
In consequence, we will assume that the operator $S_W$ can be dropped in \eqref{DilatonSoftTheorem}, and this turns out to be consistent with our observations in the subsequent \Secref{sec:confsymgrav}.%
\footnote{\label{foot:edotq}
Note that if all reference vectors used to define the polarization vectors are chosen to be equal and such that $\epsilon_i\cdot q=0$,
 the operator $S_W$ vanishes without employing any resummation. The same applies to the operator $S_V$.}

 %%%%%%%%%%%%%%%%%%
 
 \paragraph{The Operator $\op{S}_V$.}
 The operator $\op{S}_V$ comprises 
 only derivatives acting on the polarization vectors $\e_i$, and hence the quadratic dependence of $M_n^{{\a'}}$ on $\e_i$  can be exploited to rewrite these terms more transparently as
 \begin{align}
 \sum_{i=1}^{n} \frac{q_\rho q^\sigma}{ 2 k_i\cdot q} 
 \left(      
 S_i^{ \rho \mu  }  S_{i, {\mu \sigma } }
  +   d     \e_{i}^{\rho} 
\partial_{\e_{i }}^{\sigma}
\right )
 &
   \ddelta(P)\left (\e_i^\alpha\e_i^\beta  M_{n,i, \alpha \beta}^{{\a'}}\right ) 
   \label{eq:ExplicitS2}
   \\
   &= 
   \ddelta(P)\sum_{i=1}^n \frac{
q^{\alpha}q^{\beta} (\epsilon_i \cdot{\epsilon}_i) + \eta^{\alpha \beta} (q \cdot \epsilon_i)^2%
}{ k_i\cdot q}  M_{n,i, \alpha \beta}^{{\a'}}.\nonumber%
\end{align}%
\begin{figure}
 \captionsetup[subfloat]{farskip=5pt,captionskip=5pt}
\begin{center}
 \subfloat[\label{fig:soft1}]{
 \includegraphicsbox{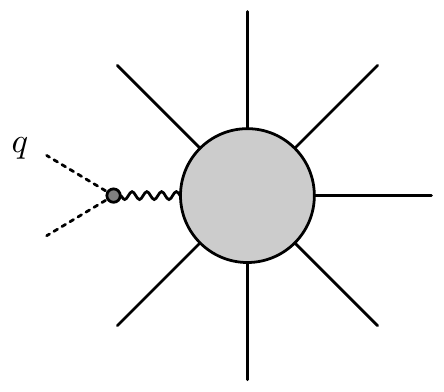}
 }
\hspace{2cm}
 \subfloat[\label{fig:soft2}]{
\includegraphicsbox{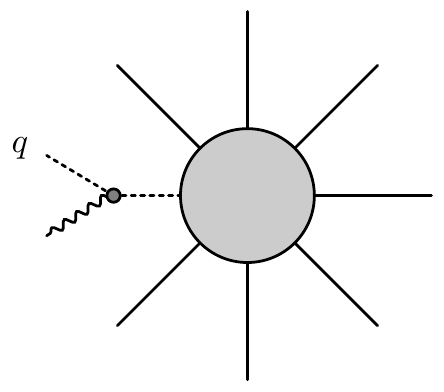}
}
\caption{Diagrammatic interpretation of the terms in  \protect\eqref{eq:ExplicitS2}. Dilatons are represented by dashed lines while gravitons correspond to wavy lines. The soft particle carries momentum $q$. 
}%
\end{center}%
\end{figure}%
In field theory, we  interpret the above terms as on-shell emissions from external states (see also~\cite{DiVecchia:2016amo}).
Since for a graviton we have $\e_\text{g} \cdot \e_\text{g} =0$ and for a dilaton $\e_i^\mu \e_i^\nu \sim \eta^{\mu \nu}$,
the first term corresponds to the decay of an internal graviton to two external dilatons, one of them soft (see \Figref{fig:soft1}). 
The second term on the other hand corresponds to the emission of a hard graviton from the soft dilaton leg (see \Figref{fig:soft2}). 
Thus, depending on whether the $i$th state is a graviton or dilaton, only one of these terms contributes. 
For clarity we consider from here on the case where all $n$ hard states are gravitons.
In that case the soft theorem in \Eq{DilatonSoftTheorem} can, as just argued, be rewritten in the form
\ea{
\label{DilatonSoftTheorem2}
\mathcal{M}_{n+ \phi}^{\alpha'} (k_1, \ldots, k_n, q)=& \sfrac{\kappa_d}{\sqrt{d-2}}\Big[\ddelta(P) 
S_\delta
+ 
\big(q\cdot \partial_P \ddelta(P)\big)S_{\delta'}\Big]
\widetilde M_n^{{\a'}} 
\\
&
+\sfrac{\kappa_d}{\sqrt{d-2}}\sum_{i=1}^n \frac{(q \cdot \epsilon_i)^2 
}{ k_i\cdot q}   \ddelta(P){M}_{n}^{\alpha'}(k_1, \ldots, \phi(k_i), \ldots, k_n)
 + {\cal{O}} (q^2 ) \, ,\nonumber
}
where in the second line we sum over $n$-point amplitudes with the $i$th state being a dilaton rather than a graviton.
 This expression points at a peculiarity: 
 In the field theory limit, tree-level scattering amplitudes of gravitons with an odd number of dilatons vanish.
The reason for this can be understood from the low-energy action of the string,
which shows (in the Einstein frame) that the dilaton couples only quadratically to gravitons:
\ea{
S_\text{low-energy} = \frac{1}{2 \kappa_d^2} \int d^d x \sqrt{- g} \left[ R - \frac{1}{2} g^{\mu \nu} \partial_{\mu} {{\phi}} \partial_{\nu} {{\phi}}  - \frac{1}{12} {\rm e}^{-\sqrt{\frac{8}{d-2}} {{\phi}} }  \left (\partial_{[\mu} B_{\nu \rho]} \right )^2 \right] \ .
\label{Action}
}
Since also the $B$-field couples only quadratically to the graviton, scattering processes involving only gravitons and dilatons do not involve virtual $B$-fields at tree level.

Translating this $Z_2$ symmetry to generic tree-level scattering processes involving dilatons and gravitons, we conlcude that only amplitudes with an even number of external dilatons are non-zero.
Furthermore, at tree level pure graviton scattering is completely described by the Einstein--Hilbert term in the action.%
\footnote{In four dimensions, these statements hold even at loop level as a consequence of $U(1)$ charge conservation, because the dilaton forms a complex $U(1)$ multiplet together with the $B$-field which in four dimensions is dual to a pseudo-scalar field.}
Therefore, taking the limit $\a' \to 0$ of \Eq{DilatonSoftTheorem2} the contribution from the  operator $\op{S}_V$ vanishes, and we end up with
the consistency condition
\ea{
0
=
\Big[\ddelta(P) 
(2-D_{\Delta=0}+q_\mu K_{\Delta=0}^\mu)
+ 
\big(q\cdot \partial_P \ddelta(P)\big)(2-D_{\Delta=0})
\Big]
\lim_{\a' \to 0} \widetilde M_n^{{\a'}} + \Ord(q^2)\, ,
\label{main}
}
where it should be understood that all $\varepsilon_i^{\mu \nu} = \e_i^\mu \e_i^\nu $ are taken to be symmetric, traceless polarization tensors corresponding to $n$ gravitons. 
This line of arguments applies, however, also to the case where the $n$ hard states comprise both gravitons and an \emph{even} number of dilatons, and thus tracelessness of $\varepsilon_i^{\mu \nu}$ is not necessary for an even subset of the set of $n$ polarization tensors.

%%%%%%%%%

\paragraph{Consequences for Graviton Scattering.}

The obstacles for intepreting \eqref{main} in terms of field theory scattering amplitudes are the facts that the formal expression $\widetilde M_n$ is not multiplied by the momentum $\delta$-function, and moreover that the equation includes a derivative of this $\delta$-function at order $q^1$ (cf.\ the discussion around \eqref{softgraviton}).

There is, however, an immediate corollary of \Eq{main}: 
At leading order $q^0$, this equation reduces to
\ea{
\ddelta(P)
\left[  
2 - \op{D}_{\Delta=0}  \right ]
\lim_{\a' \to 0} \widetilde M_n^{{\a'}}
= 
\ddelta(P)\left[  
2 - \op{D}_{\Delta=0}  \right ]
 M_n = 0,
\label{homogeneity}
}
where $ M_n$ is the stripped $n$-point graviton amplitude in field theory. To obtain the stripped amplitude, the commutator \Eq{commDdelta} was tacitly, but importantly, used. 
This corollary states that  for any number of external states $n$, field theory tree-level graviton amplitudes are homogeneous functions of degree 2 in the momenta. This is indeed a well-known fact. In Ref.~\cite{DiVecchia:2015jaq} this consistency condition was also discussed from a string theory perspective.

At subleading order, the above equation \eqref{main} reduces to
\begin{equation}
0
=
\Big[\ddelta(P) 
q_\mu K_{\Delta=0}^\mu
+ 
\big(q\cdot \partial_P \ddelta(P)\big)(2-D_{\Delta=0})
\Big]
\lim_{\a' \to 0} \widetilde M_n^{{\a'}}\, .
 \label{problem}
\end{equation}
In order to turn $\widetilde M_n^{\a'}$ into the stripped amplitude, we pull the momentum conserving $\delta$-function through the special conformal generator using the commutation relation \eqref{comOS} modulo Lorentz invariance of $\widetilde M_n^{{\a'}}$:
\begin{equation}
0
=
\Big[q_\mu \op{K}_{\Delta=0}^\mu \ddelta(P) +\big(q\cdot \partial_P  \ddelta(P)\big)  (2-d)
\Big]
\lim_{\a' \to 0} \widetilde M_n^{{\a'}}\, .
\end{equation}
We may rewrite this in the following suggestive way:
\begin{align}
0
=
&\Big[q_\mu \op{K}_{\Delta=0}^\mu \ddelta(P) -\sfrac{d-2}{n} \sum_{i=1}^n \big(q\cdot\partial_{k_i}  \ddelta(P)\big)
\Big]
\lim_{\a' \to 0} \widetilde M_n^{{\a'}}\, 
\nonumber\\
=
&q_\mu \op{K}_{\Delta=\frac{d-2}{n}}^\mu \ddelta(P)
\lim_{\a' \to 0} \widetilde M_n^{{\a'}}
-\ddelta(P) \sfrac{d-2}{n} \sum_{i=1}^n q\cdot \partial_{k_i}\lim_{\a' \to 0} \widetilde M_n^{{\a'}},
\end{align}
where in the second line we introduce a nontrivial scaling parameter 
\begin{equation}
\Delta=\frac{d-2}{n}.
\end{equation}
Hence, assuming that a representation of $\widetilde M_n^{\alpha'}$ exists for which 
\begin{equation}
\ddelta(P) {\Delta} \sum_{i=1}^n q\cdot \partial_{k_i}\lim_{\a' \to 0} \widetilde M_n^{{\a'}}=0,
\label{eq:symrep}
\end{equation}
we arrive at the statement
\begin{equation}
 \op{K}_{\Delta}^\mu\,  \mathcal{M}_n = 0 \,,
\qquad
\text{for \quad}\Delta =\frac{d-2}{n}.
\label{hypothesis}
\end{equation}
Here,  $\mathcal{M}_n=\ddelta(P) M_n$ is the $n$-graviton field theory scattering amplitude. We may translate this into an invariance statement for the $\delta$-stripped amplitude by noting that for $\Delta=\frac{d-2}{n}$ we have
\ea{
 \op{K}_{\Delta}^\mu\,  \mathcal{M}_n =\ddelta(P) \op{K}_{\Delta}^\mu M_n 
 +
\big( \partial_{P_\mu}\ddelta(P)\big)\big(d-\op{D}_{\Delta}\big)M_n 
 = \ddelta(P) \op{K}_{\Delta}^\mu M_n \,,
}
where the corollary \eqref{homogeneity} was used. Hence, we find that $\op{K}_{\Delta=\frac{d-2}{n}}^\mu$ also annihilates the $\delta$-stripped graviton amplitude:
\begin{equation}
\ddelta(P) \op{K}_{\Delta}^\mu\,  M_n = 0 \,,
\qquad
\text{for \quad}\Delta =\frac{d-2}{n}.
\label{strippedhypothesis}
\end{equation}
It follows that all generators of the conformal algebra annihilate the tree-level graviton amplitude $\mathcal{M}_n$ provided we are working on a representation of $\widetilde M_n^{\alpha'}$ which obeys \eqref{eq:symrep}. In \Appref{threepointresolution} our explicitly derived result for $\widetilde M_3^{\alpha'}$ indeed takes a form that obeys \eqref{eq:symrep}.
 In the following section we will argue that, at least for the $\delta$-stripped field theory amplitude $M_n$, such a representation can be obtained by making the permutation symmetry of graviton scattering manifest.

%%%%%%%%%%%%%%%%%%%%%%%%%%%%%%%%%%%%

\section{Conformal Symmetry of Graviton Amplitudes}
\label{sec:confsymgrav}
In this section we study the relations derived in the previous section on explicit field theory amplitudes.
We would like to understand when \eqref{eq:symrep} and \eqref{strippedhypothesis} are satisfied.
Our study of conformal transformations of Yang--Mills amplitudes in momentum space exposed the necessity to manifest physical properties of the stripped amplitude, namely cyclic permutation and reversal symmetry, in order to observe the invariance properties of amplitudes in four dimensions. The analogous property for graviton amplitudes is Bose or full permutation symmetry. 
The importance of manifesting this symmetry was also recently noticed in~\cite{McGady:2014lqa,Boels:2017gyc}.

Interestingly, it turns out that by manifesting full permutation symmetry of the stripped graviton amplitudes, the identity \eqref{eq:symrep} as well as the implication of special conformal symmetry \eqref{strippedhypothesis} are satisfied --- at least for three, four, five, and six external gravitons. This is reminiscent of the Yang--Mills case. 
Additionally, we observe that the ambiguities arising from stripping off the $\delta$-function become irrelevant when acting on the manifestly permutation symmetric amplitude.
We will detail these results in the following.

\paragraph{Constructing Graviton Amplitudes from Yang--Mills Amplitudes.}
Tree-level graviton amplitudes can be constructed from Yang--Mills amplitudes in various ways.
One way is through the so-called KLT relations~\cite{Kawai:1985xq}, which relate closed string amplitudes to a double-copy of open string amplitudes through a momentum kernel with a well-defined field theory limit (for a brief review on the field theory relations, see e.g.\ \cite{Bern:1998ug,Bern:1998sv}). 
In this work, the relations at three, four, five and six points will be used, to be precise:
\ea{
M_3 (1,2,3) &= i A_3(1,2,3)\bar{A}_3(1,2,3) \, , \\[2mm]
M_4 (1,2,3,4) &= -i s_{12} A_4(1,2,3,4)\bar{A}_4(1,2,4,3) \, , \\[2mm]
M_5(1,2,3,4,5)
& = i s_{12} s_{34}  A_5(1,2,3,4,5) \bar{A}_5(2,1,4,3,5) 
\nonumber \\
&\quad + i s_{13}s_{24} A_5(1,3,2,4,5) \bar{A}_5(3,1,4,2,5)  \, , \\[2mm]
M_6(1,2,3,4,5,6) 
& = - i s_{12} s_{45}  A_6(1,2,3,4,5,6) 
    \Big [ s_{35} \bar{A}_6(2,1,5,3,4,6) 
    \nonumber \\
& \quad      
   + (s_{34} + s_{35}) \bar{A}_6(2,1,5,4,3,6) \Big] 
 + \mathcal{P}(2,3,4) \, .
}
Here $s_{ij} = 2 k_i \cdot k_j$, and $A_n$ and $\bar{A}_n$ are the color-decomposed partial amplitudes as defined in \Eq{colordecomposition}, with the bar on $\bar{A}_n$ indicating that the polarization vectors can be distinguished, i.e.\ in $\bar{A}_n$ we take
 $ \e_i \to \bar{\e}_i$. Being interested in graviton amplitudes, however, we need to set $\bar{\e}_i \to \e_i$, or simply $\bar{A}_n \to A_n$.  The term $\mathcal{P}(2,3,4)$ indicates a sum over all permutations of the indices $2,3$ and $4$.
The graviton amplitudes are then given by
\ea{
\mathcal{M}_n = \ddelta(P) \left (\frac{\kappa_d}{2}\right)^{n-2} M_n \, ,
}
with $\kappa_d$ denoting the $d$-dimensional gravitational constant. 
As remarked, the polarization vectors in the two copies of Yang--Mills amplitudes are identified with the graviton polarization tensor by $\e^\mu \e^\nu \equiv \varepsilon_g^{\mu \nu}$ with $\e \cdot \e = 0$. The stripped amplitude $M_n$ is obtained by implementing the relations \Eq{MomPoincare}.
The amplitude is fully permutation symmetric, i.e.\ we have
\ea{
\mathcal{M}_n (1, 2, \ldots, n) = \mathcal{M}_n ( \mathcal{P}(1,2, \ldots, n) ) \, ,
}
as a consequence of Bose symmetry. We stress that this identity is generically not manifest, but holds modulo momentum conservation and on-shell conditions. For our purposes, however, it is useful to make this symmetry manifest at the level of the stripped amplitude by explicitly permutation symmetrizing it. The notation $\mathcal{P}_n[M_n]$ will be used to indicate this manipulation:
\begin{equation}
\mathcal{P}_n[f(k_1,\epsilon_1,\dots,k_n,\epsilon_n)]=\frac{1}{n!}\sum_{{\cal P}(1, \ldots, n)}f(k_1,\epsilon_1,\dots,k_n,\epsilon_n).
\end{equation}
Here $f$ denotes a function of $n$ momenta and polarization vectors.%
\footnote{Note that  
for $f=M_n|_{k_n=-\sum_{i=1}^{n-1}k_i}$
the dependence of $f$ on $k_n$ will be trivial before symmetrization. 
The symmetrization $\mathcal{P}_n$ in $n$ variables, however, reintroduces $k_n$. }
As for $\mathcal{C}_n$ in the Yang--Mills case \Eq{eq:CnAnequalsAn}, we have of course that 
\ea{
\mathcal{P}_n [M_n] = M_n \, ,
}
where by equality we mean on resolving the Poincaré constraints as prescribed by \Eq{MomPoincare} (as in all previous discussions).

\paragraph{Conformal Invariance of Graviton Amplitudes.}
We would like to establish explicitly whether tree-level graviton amplitudes can be considered conformally invariant in the sense described in the previous section.
Let us first point out some general features. As in~\Eq{Poincare},
Poincare invariance constrains the action of $\op{K}_\Delta^\mu$ on $M_n$ to the form
\ea{
\op{K}_\Delta^\mu M_n = 
{\sum_{i=1}^n} \epsilon_i^\mu F_i + {\sum_{i=1}^n} k_i^\mu G_i \, .
\label{eq:strucgravity}
}
In all cases to be discussed below, we curiously find that
\ea{
\ddelta(P) \sum_{i=1}^n k_i^\mu G_i  \propto \Delta \, .
\label{Gigrav}
}
Hence, for $\Delta =0$ the second sum in \eqref{eq:strucgravity} vanishes, reminiscent of what happens for Yang--Mills amplitudes for $\Delta=1$, cf.\ \Eq{Gi}.
While in both cases we lack a rigorous understanding of this feature, we notice that $\Delta=0$ is the classical scaling dimension of a graviton in $d=2$, where gravity is known to be conformal.
Rephrasing the above, we thus have generically%
\footnote{Note that if all reference vectors used to define the polarization vectors are chosen to be equal and such that $\e_i\cdot q=0$ holds, we find
$
q\cdot\op{K}_{\Delta=0} M_n = 0 .
$
See Footnote~\ref{foot:edotq} in this context.
}
\ea{
\op{K}_{\Delta=0}^\mu M_n = 
{\sum_{i=1}^n} \epsilon_i^\mu F_i \neq 0\, .
}
 This similarity to the $d=4$ Yang--Mills case (cf.\ e.g.\ \eqref{eq:K1neq0}) suggests that the above nontrivial functions $F_i$ arise from a deficiency of preserving permutation symmetry.
In fact, it turns out that in the studied examples up to (and including) six points we have
\ea{
\op{K}_{\Delta}^\mu \, \mathcal{P}_n [M_n] = 0 \, ,
\label{result1}
}
for any value of $\Delta$. 
In particular, this implies%
\footnote{On the amplitude which only depends on Lorentz invariants, the vanishing of the summed momentum derivative, i.e.\ $\sum_{i=1}^n \partial_{k_i}^\mu \,  \mathcal{P}_n [M_n] = 0 $, can be alternatively stated in terms of the two conditions
$\sum_{i}\partial_{s_{ij}} \mathcal{P}_n [M_n]=0$ and $\sum_{i} \partial_{{\epsilon_j \cdot k_i}} \mathcal{P}_n [M_n]=0$ (cf. \Appref{App:ConformalGenerators}).
}
\begin{align}
\op{K}_{\Delta=0}^\mu \, \mathcal{P}_n [M_n] &= 0,
&\text{and}&&
\sum_{i=1}^n \partial_{k_i}^\mu \,  \mathcal{P}_n [M_n] &= 0 \, .
\label{result2}
\end{align}
Hence, $\mathcal{P}_n [M_n] $ furnishes a representation of the field theory amplitude which obeys the condition \eqref{eq:symrep} in the previous \Secref{sec:softtheorem}.
As discussed in that section, choosing the peculiar value 
\begin{equation}
\label{eq:Deltagrav}
\Delta=\frac{d-2}{n},
\end{equation}
 the above results together with (cf.\ \eqref{homogeneity})
 \begin{equation}
 \op{D}_{\Delta=\frac{d-2}{n}}\ddelta(P)\mathcal{P}_n [M_n]= \op{D}_{\Delta=\frac{d-2}{n}}\ddelta(P)M_n=0,
\end{equation}
establish our conjecture of full conformal invariance of tree-level graviton amplitudes in ordinary Einstein gravity. 
 We have verified these statements analytically and for generic spacetime dimension $d$ up to multiplicity four. For mulitplicity five and six we have checked the above conjecture \eqref{result2} by numerical evaluation of the coefficients $F_i$ and $G_i$ in \eqref{eq:strucgravity} using four-dimensional kinematical data generated by S@M~\cite{Maitre:2007jq}. These results are insensitive to the ambiguities in obtaining the stripped amplitude by resolving the Poincar\'e constraints as described in (\ref{MomPoincare}).
 The $n$-dependence of the scaling dimension $\Delta$ in \eqref{eq:Deltagrav} suggests that these results do not follow from the Lagrangian description of gravity in a straightforward manner.

\paragraph{Three-Point Example.}
At three points the KLT momentum kernel is unity, and the three-point graviton amplitude is simply given by the square of the color-ordered Yang--Mills amplitude:
\ea{
\mathcal{M}_3 \sim \ddelta(P)  A_3(1,2,3)^2 \, .
}
The stripped graviton amplitude thus reads $M_3 = A_3(1,2,3)^2$, and we again resolve Poincar\'e symmetry as prescribed by~\Eq{MomPoincare}.
Now, using the expression in \Eq{eq:A3} for $A_3$, one observes that
\ea{
\op{K}_\Delta^\mu M_3 = 4 (1+ \Delta) A_3 \left (\op{K}_\Delta^\mu \, A_3 \right ),
\label{eq:KM3}
}
which coincidentally vanishes for $\Delta = -1$, but not otherwise.
If instead of $A_3(1,2,3)^2$ we consider $\mathcal{P}_3[A_3^2]$ or $\mathcal{C}_3[A_3]^2$ (notice that the latter is also permutation symmetric), we observe that \Eq{result2} holds
for any $\Delta$.

%%%%%%%%%%%%%%%%%%%%%%%

\paragraph{Four-Point Example.}
At four points starting from the stripped amplitude $M_n$, we find by the exact same procedure as in the three-point case that \Eq{result1} and \Eq{result2} are fulfilled. We have additionally, like in the case of Yang--Mills theory, also studied the relations starting from the manifestly gauge invariant expression for $\mathcal{M}_n$ obtained by using the expression \Eq{t8expression} for $A_4$ in the KLT relations, which in terms of the stripped amplitude reads
\ea{
M_4 = 
(-i) \frac{16}{s_{12}s_{23}s_{13}} (t_{8, \mu_1 \nu_1, \ldots, \mu_4 \nu_4} k_1^{\mu_1} \cdots k_4^{\mu_4} \epsilon_1^{\nu_1} \cdots \epsilon_4^{\nu_4})^2 \, .
}
This expression has the virtue of being manifestly gauge invariant. It is additionally manifestly permutation symmetric in the labels $1,2,3$ upon replacement of $k_4 = - (k_1 + k_2 + k_4)$.
It can be readily checked using this expression that \Eq{result1} and \Eq{result2} are also  fulfilled in this special case.

In Appendix~\ref{App:2dil2grav} we provide in addition an explicit demonstration of (\ref{result1}) in the case of two-dilaton+two-graviton scattering, where this equation is also expected to hold (cf.\ the discussion below (\ref{main})).

\paragraph{Five- and Six-Point Checks.}
At five- and six-points we find again by the same procedure as detailed in the three-point case that the relations \Eq{result1} and \Eq{result2} are fulfilled. The details should by now be clear, but let us make a remark about the computational complexity in observing \Eq{result1} and \Eq{result2}:
At $n=6$, the amplitude $M_6$ contains, in expanded form and before any reduction, roughly $10^8$ terms. Permutation symmetrizing increases this number by a factor of $5!$ (since the KLT expression is already symmetric in three labels), while the action of $\op{K}_{\Delta}^\mu$ boosts the number of terms at least by two orders of magnitude, ending naively with some $\sim 10^{12}$ terms, that in the end sum to zero.

%%%%%%%%%%%%%%%%%%%%%%%%%%%%%%%%

\section{Summary and Outlook}

The aim of the present paper was to identify the consequences of the string theory soft dilaton limit for graviton scattering. We found that the indications for a conformal symmetry can indeed be turned into an invariance statement of tree-level graviton amplitudes under the full conformal algebra, though at the cost of a multiplicity dependent scaling dimension. Moreover, the formulation of the symmetry in terms of differential operators in momenta and polarization vectors, 
which does not leave the on-shell constraint surface invariant, hinges on the full symmetrization of the graviton amplitude. Dropping the symmetrization prescription, we observe similarities to the special conformal symmetry of the (unsymmetrized) Yang--Mills amplitude in four dimensions, but no full conformal symmetry.

Let us summarize our observations in some more detail.
Clearly, gluon and graviton amplitudes are both invariant under Poincar\'e symmetry. Invariance of the full amplitude (including the momentum conserving $\delta$-function) under dilatations generated by $\op{D}_\Delta$  requires to have
\begin{align}\label{eq:diladelta}
&\text{YM}: \Delta=\frac{d-4}{n}+1, 
&
&\text{Gravity}: \Delta=\frac{d-2}{n}.
\end{align}
These values for $\Delta$ become multiplicity independent in the spacetime dimensions $d=4$ and $d=2$, respecively, for which the conformal symmetry of the respective theories is well known. The most involved analysis concerns the generator $K^\mu_\Delta$ of special conformal transformations.
Poincar\'e invariance implies that the action of this generator on the stripped Yang--Mills or gravity amplitude, here collectively denoted by $A_n$, takes the form
\begin{equation}
\label{eq:Kstrucsumm}
\op{K}_\Delta^\mu A_n = 
{\sum_{i=1}^n} \epsilon_i^\mu F_i + {\sum_{i=1}^n} k_i^\mu G_i \, .
\end{equation}
In the case of Yang--Mills theory we find that $G_i=0$ for $\Delta=1$.
In the case of gravity we find $G_i=0$ for $\Delta=0$. 

The special conformal generator commutes with the momentum conserving $\delta$-function only for the values \eqref{eq:diladelta} of $\Delta$, which imply dilatation invariance of the full amplitude.

The non-vanishing of the coefficients $F_i$ in \eqref{eq:Kstrucsumm} appears to be related to the incompatibility of $K_\Delta^\mu$ with the on-shell constraints --- at least for the conformal Yang--Mills theory in 4d. We thus make the physical symmetries of the amplitudes manifest, i.e.\ cyclic/reversal or full permutation symmetry, respectively. In the case of Yang--Mills theory in $d=4$ and for $\Delta=1$, we then indeed find $F_i=0$, if the constraints from momentum conservation are resolved as prescribed by \eqref{eq:MomChoice}. We note that, somewhat unsatisfactorily, different ways of resolving these constraints do not always give this result. We attribute the latter observation to the fact that not all symmetries of the YM amplitude are manifest after cyclization and reversal. 

For the case of fully permutation symmetrized graviton amplitudes, however, this ambiguity seems to play no role and we find $F_i=0$ for any value of $\Delta$ and $d$.
Since in the gravity case manifest symmetrization leads to invariance under the special conformal generator independently of the value of $\Delta$, we may choose the scaling dimension as in \eqref{eq:diladelta}, which guarantees also dilatation invariance and a vanishing commutator of $K^\mu_\Delta$ with $\ddelta(P)$. Hence, for this multiplicity dependent choice of the scaling dimension we observe full conformal symmetry of tree-level graviton amplitudes, if the amplitudes are in a manifestly permutation symmetric form.

The observations of this paper lead to a tower of follow-up questions. The most pressing point is the meaning of the conformal properties found here. Do they represent curious coincidences or is there a deeper significance behind them? In particular, it would be important to overcome the specific symmetrization prescription employed here, which is due to the use of momentum and polarization variables. Formulating our observations in a form that manifestly preserves the on-shell constraints should distill the physical content of these statements.

Another point is whether the finding of a conformal symmetry of tree-level graviton amplitudes \emph{in any dimension} may also be transferred to the Yang--Mills case, once we manifest all symmetries of the amplitude and allow for a multiplicity dependent scaling dimension. 
If this would indeed be the case, it would emphasize the need for an interpretation of our results.

A natural question is whether the above conformal symmetry of graviton amplitudes can be deduced from the Einstein--Hilbert action. In order to approach this problem, a convenient starting point might be the simplification of this action as expressed in \cite{Cheung:2017kzx,Tomboulis:2017fim}, which employs only cubic interactions (see also~\cite{Manvelyan:2010wp,Manvelyan:2010je}), or by considering the twistor action of \cite{Adamo:2013tja}.
In fact, it should be very enlightening to make connection to twistor methods for gravity amplitudes (see \cite{Skinner:2013xp,Adamo:2013tja} and references therein). Intriguingly, these approaches were motivated by Maldacena's embedding of tree-level Einstein gravity into conformal gravity in curved space \cite{Maldacena:2011mk} (see also~\cite{Metsaev:2007fq}), as well as by Hodges' determinant expression for MHV graviton amplitudes making Bose symmetry manifest~\cite{Hodges:2012ym}.
In relation to our observations in pure Einstein gravity, it would also be interesting to understand how the conformal symmetry of amplitudes in conformal (super)gravity is realized, see e.g.~\cite{Kaku:1978ea,Berkovits:2004jj,Johansson:2017srf}. The latter analysis might help in translating our statements into spinor-helicity or twistor variables.

We speculate that BCFW recursion relations~\cite{Britto:2004ap,Britto:2005fq}, valid for tree-level graviton amplitudes~\cite{Bedford:2005yy,Cachazo:2005ca}, could provide another path towards a proof of the conformal properties presented in this paper. In fact, permutation symmetry seems also to play a pivotal role in the recursive constructibility of graviton tree-amplitudes~\cite{McGady:2014lqa}.

Recently,  tree-level gluon amplitudes were mapped to correlators on the 2d celestial sphere \cite{Pasterski:2017ylz,Schreiber:2017jsr}. Here, the two-dimensional conformal symmetry of these correlators originates from the Lorentz symmetry of the 4d gluon amplitudes. It would be interesting to understand which role the 4d conformal symmetry plays on the celestial sphere and to extend this analysis to the graviton case.

Our motivation to look for conformal properties of graviton amplitudes was largely based on a detailed analysis of the string theory soft dilaton limit derived in \cite{DiVecchia:2015oba,DiVecchia:2016amo,DiVecchia:2016szw,DiVecchia:2017gfi}. It would be important to see whether one can deduce similar consequences of soft limits for field theory graviton amplitudes at loop order, and whether the conformal generators also play a role there.

Another interesting direction would be to turn the logic of this paper around: Assuming conformal symmetry of graviton tree-amplitudes, what are the implications for single or multiple soft graviton limits? Considering the analysis of \cite{Larkoski:2014hta} concerning a similar question in the Yang--Mills case, strong constraints may be expected. 
Similarly one may wonder which implications the observed conformal invariance has on the 
KLT relations or the color-kinematics duality. 
Does it constrain the KLT kernel or the BCJ kinematic numerators?
And how can our results be seen in the new formalism of Cachazo, He and Yuan~\cite{Cachazo:2013hca} 
for Yang--Mills and gravity amplitudes?

On the other hand,  
these modern formulations of gravity amplitudes
can be taken as a motivation to search for the imprints of the conformal symmetry of gluon scattering in their gravitational double copy.
Intriguingly, we note that in the maximally supersymmetric extension of 4d Yang--Mills theory, the $\mathcal{N}=4$ super Yang--Mills model, this conformal symmetry
is in fact not only lifted to a  $\mathcal{N}=4$ superconformal one, but rather
to an infinite dimensional symmetry algebra known as the Yangian \cite{Drummond:2009fd}. The latter is in turn related to a dual superconformal symmetry \cite{Drummond:2008vq}, and persist also in supersymmetric gauge theories in $d=3,6$ and $10$ dimensions \cite{Bargheer:2010hn,Gang:2010gy,Dennen:2010dh,CaronHuot:2010rj}. In fact, these symmetry
structures are constructive, in the sense that they determine the form of the tree-level
scattering amplitudes \cite{Bargheer:2009qu,Korchemsky:2009hm}. One may thus wonder whether these extended conformal symmetries leave an imprint in their respective supergravity double copies.

We hope that the present work offers new grounds for understanding some of the many mysterious features seen in recent years in gravity and Yang--Mills theory.

%%%%%%%%%%%%%%%%%

\subsection*{Acknowledgements}

MM would like to thank Yegor Korovin for collaboration at the early stages of this work, as well as Henrik Johansson for clarifying communication on the field theory properties of dilaton scattering.
We are grateful to Niklas Beisert, Paolo Di Vecchia, Henrik Johansson, Raffaele Marotta, Tristan McLoughlin, Karapet Mkrtchyan and Oliver Schlotterer for helpful discussions and for comments on our manuscript.
The work of FL is funded by the Deutsche Forschungsgemeinschaft (DFG, German Research Foundation) -- Projektnummer 363895012.

%%%%%%%%%%%%%%%%%%%%%%%%%%%%%%%%

\appendix
\section{Conformal Generator on Stripped Amplitudes}\label{App:ConformalGenerators}

We rewrite the action of $D_\Delta$ and $K_\Delta^\mu$, given in \Eq{ConformalOperators}, on $\delta$-stripped amplitudes in terms of differential operators of the kinematic invariants:
\ea{
s_{ij} = p_i \cdot p_j \, , \quad w_{ij} = \epsilon_i \cdot p_j \, , \quad e_{ij} = \epsilon_i \cdot \epsilon_j
\, . 
\label{kininv}
}
Notice that $s_{ij}$ is here different by a factor of two from the main text. 
The $n$-point stripped amplitude can be considered a function of these invariants as follows:
\ea{
A_n = A_n(s_{i<j, j}, w_{i\neq j,j}, e_{i < j, j} ) \quad \forall i, j = 1, \ldots, n \, .
}
Notice that we are assuming the amplitude to be a function of on-shell variables only, i.e. $s_{ii} = w_{ii} = e_{ii}=0$. The latter assumption $e_{ii} = 0$ is a property of the graviton, but does not change the more general KLT amplitudes, involving gravitons, dilatons and B-fields, since $e_{ii}$ do not enter there.

We introduce the differential operators
\ea{
\partial_{X_{ij}} \equiv \frac{\partial}{\partial X_{ij}}  \, \quad \text{for} \quad X = \{s, w, e\}\, ,
}
and we identify
\ea{
\partial_{s_{ij}} = \partial_{s_{ji}}  \quad \text{and} \quad \partial_{e_{ij}} = \partial_{e_{ji}} 
\, ,
}
similar to the property of $s_{ij}$ and $e_{ij}$. This is important for taking into account the correct counting.%%%%%
%%%% SECRET SOURCE NOTE: 
%%%%In Mathematica, the property $X_{ij}=X_{ji}$ can be implemented using Attributes[X] = Orderless

With these ingredients it is straightforward to derive the following relations considering \Eq{ConformalOperators} and using the chain rule:
\ea{
D_\Delta A_n = &\sum_{i, j\neq i} \Big [\Delta + s_{ij} \partial_{s_{ij}} + w_{ji} \partial_{w_{ji}} \Big ] A_n,
\\[2mm]
K_\Delta^\mu A_n = &\sum_{i, j\neq i, l\neq i}
\Big [ 
{\textstyle \frac{1}{2}} p_i^{\mu } \left(e_{jl} \partial_{w_{ji}} \partial_{w_{li}}+2 w_{jl}
   \partial_{w_{ji}} \partial_{s_{il}}+s_{jl} \partial_{s_{ij}} \partial_{s_{il}}\right)
      \nonumber \\
&
 -p_j^{\mu }
   \left(e_{il} \partial_{w_{ij}} \partial_{w_{li}}+w_{il} \partial_{w_{ij}}
   \partial_{s_{il}}+w_{li} \partial_{s_{ij}}
   \partial_{w_{li}}+s_{il} \partial_{s_{ij}} \partial_{s_{il}}\right)
  \nonumber \\ 
 &
+ \epsilon _i^{\mu } \left(w_{lj} \partial_{e_{il}} \partial_{s_{ij}}+e_{jl} \partial_{e_{il}}
   \partial_{w_{ji}}+s_{jl} \partial_{s_{ij}} \partial_{w_{il}}+w_{jl} \partial_{w_{ji}}
   \partial_{w_{il}}\right)
       \nonumber \\
&
   -\epsilon _j^{\mu } \left(w_{il} \partial_{e_{ij}}
   \partial_{s_{il}}+e_{il} \partial_{e_{ij}} \partial_{w_{li}}+s_{il} \partial_{w_{ji}}
   \partial_{s_{il}}+w_{li} \partial_{w_{ji}} \partial_{w_{li}}\right)  
\Big ] A_n
\nonumber \\
&- \sum_{i, j\neq i} \Delta  \Big [  p_j^\mu \partial_{s_{ij}} + \epsilon_j^\mu \partial_{w_{ji}} \Big ] A_n
\, .
   }

\section{Computation of $\widetilde M_3^{\a'}$ 
}
\label{threepointresolution}

In this appendix we study $\widetilde M_n^{\a'}$ entering in \Eq{softtheorem}, where overall momentum conservation involves $n+1$ light-like momenta, for the simplest case of $n=3$.

With $n+1$-point momentum conservation, instead of $n$-point momentum conservation, $\widetilde M_n^{\a'}$ depends in principle also on how the measure $dV_{abc}$ in \Eq{Mn} is fixed (the $\mathrm{SL}(2,\mathbb{C})$ M\" obius symmetry). 
This means that \Eq{softtheorem} has the potential \emph{technical} problem of not being well-defined for a generic choice of $dV_{abc}$. 
On the other hand, \Eq{softtheorem} says that there \emph{must be a way} to fix $dV_{abc}$ that makes the right-hand side of the expression well-defined through order $q$, since the
 the left-hand side can equally well be calculated by first integrating and then expanding in $q$.
 To understand this better, we consider the simplest case of $n=3$ with momentum conservation $P+q = k_1 +k_2 + k_3 + q= 0$, with $k_i^2 = q^2 = 0$.

For brevity, denote the integrand in \Eq{Mn} by $I_n = I_n^L I_n^R$, where $L$ and $R$ indicate the holomorphic, respectively antiholomorphic parts, and introduce the compact notation $\Theta_i^\mu = \theta_i \epsilon_i^\mu$, as well as the rescaling $K_i^\mu = \sqrt{\frac{\alpha'}{2}} k_i^\mu$ and $Q^\mu = \sqrt{\frac{\alpha'}{2}} q^\mu$, thus
\ea{
I_n^L =
\prod_{i<j}^n (z_i - z_j )^{ K_i K_j}  
\exp \left[ \sum_{i< j}^n \frac{\Theta_i \cdot \Theta_j}{(z_i - z_j)^2}+  \sum_{i \neq j}^n \frac{ \Theta_i \tcdot K_j }{z_i - z_j}  \right]  \, . 
}
For $n=3$ there is no integration to be done, because all moduli are completely fixed by the M\"obius symmetry. However, this is in principle only well defined for $P=\sum_{i=1}^n k_i = 0$. Here we instead study the integration under the constraint $P+q=0$. To see the problem and possible resolution,
let us study the textbook method of doing this calculation. The integration is fixed by
\ea{
dV_{123} = \frac{d^2 z_1 d^2 z_2 d^2 z_3}{ |z_1-z_2|^2  |z_2-z_3|^2  |z_3-z_1|^2} \, .
}
The standard choice for fixing the three points is at $z = 0, 1, \infty$. 
Since there is no integration to be done, we can just as well consider only one holomorphic part (which is equivalent to studying the open-string three-point case) and then multiply with the antiholomorphic part in the end. After performing the Gra{\ss}mann integration, the holomorphic part reads (denoting $\e_i \cdot \e_j \equiv e_{ij}$):
\ea{
& \int \frac{\prod_{i=1}^3 d z_i }{dV_{123}^L} \int \prod_{i=1}^n d\theta_i  \ I_n^L =
 \nonumber \\
&(z_2-z_1)^{1+K_1K_2}  (z_2-z_3)^{1+K_2K_3}  (z_3-z_1)^{1+K_1K_3}
\Bigg [
\frac{e_{12}}{(z_1-z_2)^2}\left ( \frac{\e_3\tcdot K_1}{z_3-z_1}+\frac{\e_3\tcdot K_2}{z_3-z_2} \right )
\nonumber \\
&
+
\frac{e_{13}}{(z_1-z_3)^2}\left ( \frac{\e_2\tcdot K_1}{z_2-z_1}+\frac{\e_2\tcdot K_3}{z_2-z_3} \right )
+
\frac{e_{23}}{(z_2-z_3)^2}\left ( \frac{\e_1\tcdot K_2}{z_1-z_2}+\frac{\e_1\tcdot K_3}{z_1-z_3} \right )
\nonumber \\
&
\left ( \frac{\e_1\tcdot K_2}{z_1-z_2}+\frac{\e_1\tcdot K_3}{z_1-z_3} \right )
\left ( \frac{\e_2\tcdot K_1}{z_2-z_1}+\frac{\e_2\tcdot K_3}{z_2-z_3} \right )
\left ( \frac{\e_3\tcdot K_1}{z_3-z_1}+\frac{\e_3\tcdot K_2}{z_3-z_2} \right )
\Bigg] \, .
}
Choosing $z_1=0, \ z_2 = 1, \ z_3 = z \to \infty$, we get
\ea{
=
&(1-{z})^{1+K_2K_3}  ({z})^{1+K_1K_3}
\Bigg [
\frac{e_{12}}{(1)^2}\left ( \frac{\e_3\tcdot K_1}{{z}}-\frac{\e_3\tcdot K_2}{1-{z}} \right )
\nonumber \\
&
+
\frac{e_{13}}{({z})^2}\left ( \frac{\e_2\tcdot K_1}{1}+\frac{\e_2\tcdot K_3}{1-{z}} \right )
+
\frac{e_{23}}{(1-{z})^2}\left (- \frac{\e_1\tcdot K_2}{1}-\frac{\e_1\tcdot K_3}{{z}} \right )
\nonumber \\
&
\left ( -\frac{\e_1\tcdot K_2}{1}- \frac{\e_1\tcdot K_3}{{z}} \right )
\left ( \frac{\e_2\tcdot K_1}{1}+\frac{\e_2\tcdot K_3}{1-{z}} \right )
\left ( \frac{\e_3\tcdot K_1}{{z}}-\frac{\e_3\tcdot K_2}{1-{z}} \right )
\Bigg]\, .
}
Expanding and neglecting terms of $\Ord(1/{z})$ we get
 \ea{
=
&(1-{z})^{K_2K_3}  ({z})^{K_1K_3}
\Bigg [
e_{12}(\e_3\tcdot K_1)-{z} \, e_{12} \, \e_3\cdot( K_1 + K_2)
\nonumber \\
&
+
\frac{1-{z}}{{z}} \  e_{13} \left ( \e_2\tcdot K_1\right )
-\frac{{z}}{1-{z}}\ e_{23} \left ( \e_1\tcdot K_2 \right )
\nonumber \\
&
-(1-{z})\left (\e_1\tcdot K_2\right)\left (\e_2\tcdot K_1\right)\left (\e_3\tcdot K_1\right)
+{z}\left (\e_1\tcdot K_2\right)\left (\e_2\tcdot K_1\right)\left (\e_3\tcdot K_2\right)
\nonumber \\
&
- \left (\e_1\tcdot K_2\right)\left (\e_2\tcdot K_3\right)\left (\e_3\tcdot K_1\right)
+\frac{{z}}{1-{z}}\left (\e_1\tcdot K_2\right)\left (\e_2\tcdot K_3\right)\left (\e_3\tcdot K_2\right)
\nonumber \\
&
- \frac{1-{z}}{{z}}  \left(\e_1\tcdot K_3\right) \left (\e_2\tcdot K_1\right) \left (\e_3\tcdot K_1\right)
+ \left (\e_1\tcdot K_3\right)\left (\e_2\tcdot K_1\right)\left (\e_3\tcdot K_2\right)
+ \Ord(1/{z}) \Bigg] 
\nonumber \\
=&
e^{ (K_1+K_2)\tcdot K_3 \ \ln ({z}) }
\Bigg [
e_{12}(\e_3\tcdot K_1)-{z} \, e_{12} \, \e_3\cdot( K_1 + K_2)
\nonumber \\
&
-  e_{13} \left ( \e_2\tcdot K_1\right )
+ e_{23} \left ( \e_1\tcdot K_2 \right )
-\left (\e_1\tcdot K_2\right)\left (\e_2\tcdot K_1\right)\left (\e_3\tcdot K_1\right)
\nonumber \\
&
+{z}\left (\e_1\tcdot K_2\right)\left (\e_2\tcdot K_1\right)\, \e_3\cdot \left (K_1+ K_2\right)
- \left (\e_1\tcdot K_2\right)\left (\e_2\tcdot K_3\right)\, \e_3\cdot \left (K_1+ K_2\right)
\nonumber \\
&
+ \left(\e_1\tcdot K_3\right) \left (\e_2\tcdot K_1\right)\, \e_3\cdot \left (K_1+ K_2\right)\Bigg] 
+ \Ord(1/{z}) \, .
}
Clearly this is not well-defined at ${z} = \infty$ if $K_1 + K_2+ K_3 \neq 0$.
Let us impose momentum conservation $K_1 + K_2+ K_3 = - Q$ (as well as on shell conditions) to make this more clear:
\ea{
&=
z^{- Q \tcdot K_3 }
\Bigg [
e_{12}(\e_3\tcdot K_1) + e_{23} \left ( \e_1\tcdot K_2 \right ) +  e_{13} \left ( \e_2\tcdot K_3\right )
+\left (\e_1\tcdot K_2\right)\left (\e_2\tcdot K_3\right)\left (\e_3\tcdot K_1\right)
\nonumber \\
&
+ e_{13} \left (\e_2\tcdot Q \right )
+\left (\e_1\tcdot K_2\right)\left (\e_3\tcdot K_1\right)\left (\e_2\tcdot Q \right )
+ 
\left (\e_3\tcdot Q \right )\left (
\left (\e_1\tcdot K_2\right)\left (\e_2\tcdot K_3\right)
- \left(\e_1\tcdot K_3\right) \left (\e_2\tcdot K_1\right) \right )
\nonumber \\
&
+ {z} \, \left (\e_3\tcdot Q \right ) \left ( 
\, e_{12} 
-\left (\e_1\tcdot K_2\right)\left (\e_2\tcdot K_1\right)\right )
\Bigg] 
+ \Ord(1/{z}) \, .
\label{integrated}
 }
 Here we used momentum conservation to get the cyclic permutation symmetric form in the first line, which is nothing but the three-point open string expression in the bosonic string.
 The second and third line shows the deviation for nonzero $Q$.

Consider the first line of \Eq{integrated} only. If we cyclically symmetrize $z_1, z_2, z_3$ over $\{0,1, \infty\}$ when fixing the M\"obius symmetry, we would, because of momentum conservation and $Q^2 = 0$, obtain
\ea{
z^{- Q\cdot (K_1 + K_2 + K_3) } A_3^{\a'} (1,2,3)
= z^0 A_3^{\a'} (1,2,3) = A_3^{\a'}(1,2,3) \, ,
}
where $A_3^{\a'}$ is equal to
\ea{
A_3^{\a'} (1,2,3) = e_{12}(\e_3\tcdot K_1) + e_{23} \left ( \e_1\tcdot K_2 \right ) +  e_{13} \left ( \e_2\tcdot K_3\right )
+\left (\e_1\tcdot K_2\right)\left (\e_2\tcdot K_3\right)\left (\e_3\tcdot K_1\right) .
\label{A3}
}
This shows that for well-definedness at $z=\infty$ it is here necessary to keep cyclic permutation symmetry, when performing the integration by fixing the points.

Next, let us consider the terms with the factor $z$ in the third line of \Eq{integrated} aftercyclically permutation symmetrizing the result:
\ea{
 &z^{1-Q\tcdot K_3} \left (\e_3\tcdot Q \right ) \left [ e_{12} - \left (\e_1\tcdot K_2\right)\left (\e_2\tcdot K_1\right)\right ]
 \nonumber \\
 &+
  z^{1-Q\tcdot K_2} \left (\e_2\tcdot Q \right ) \left [ e_{31} - \left (\e_3\tcdot K_1\right)\left (\e_1\tcdot K_3\right)\right ]
  \nonumber \\
  &
   +
  z^{1-Q\tcdot K_1} \left (\e_1\tcdot Q \right ) \left [ e_{23} - \left (\e_2\tcdot K_3\right)\left (\e_3\tcdot K_2\right)\right ]\, .
  }
$A_3$ should on top of cyclic symmetry also satisfy the \emph{reversal symmetry}
\ea{
A_3^{\a'}(1,2,3) = - A_3^{\a'} (3,2,1) \, .
} 
We can ensure this by also imposing this reversal symmetry, when fixing the M\"obius symmetry. This would send the above terms to
\ea{
 &z^{1-Q\tcdot K_3} \left (\e_3\tcdot Q \right ) \left [ e_{12} - \left (\e_1\tcdot K_2\right)\left (\e_2\tcdot K_1\right)\right ]
 -
 z^{1-Q\tcdot K_1} \left (\e_1\tcdot Q \right ) \left [ e_{32} - \left (\e_3\tcdot K_2\right)\left (\e_2\tcdot K_3\right)\right ]
 \nonumber \\
 &+
  z^{1-Q\tcdot K_2} \left (\e_2\tcdot Q \right ) \left [ e_{31} - \left (\e_3\tcdot K_1\right)\left (\e_1\tcdot K_3\right)\right ]
  -
  z^{1-Q\tcdot K_2} \left (\e_2\tcdot Q \right ) \left [ e_{13} - \left (\e_1\tcdot K_3\right)\left (\e_3\tcdot K_1\right)\right ]
  \nonumber \\
  &
   +
  z^{1-Q\tcdot K_1} \left (\e_1\tcdot Q \right ) \left [ e_{23} - \left (\e_2\tcdot K_3\right)\left (\e_3\tcdot K_2\right)\right ]
  -
    z^{1-Q\tcdot K_3} \left (\e_3\tcdot Q \right ) \left [ e_{21} - \left (\e_2\tcdot K_1\right)\left (\e_1\tcdot K_2\right)\right ]
    \nonumber \\
    &= \, 0 \ \, .
}
 We have thus shown that the integration is well-defined when the momentum is conserved up to a light-like deformation, here parametrized by $Q$, by preserving the symmetry properties of the integrand when fixing the M\"obius symmetry.
 
 The expression \Eq{integrated} contains additional terms depending on $Q$; i.e.\ all terms in the second line of \Eq{integrated}.
 In the end, we want only to consider the expression up to $\Ord(q^2)$, so
 we can simply set the prefactor $z^{-Q\tcdot K_3} = 1 + \Ord(q)$ for those terms.
 Then one finds that the remaining terms all vanish up to $\Ord(q^2)$.
 This is easy to see for the term of order ${\alpha'}^0$, which after cyclic and reversal symmetrization reads
 \ea{
 e_{13} \left (\e_2\tcdot Q \right )+e_{12} \left (\e_3\tcdot Q \right )+e_{23} \left (\e_1\tcdot Q \right )
 - e_{31} \left (\e_2\tcdot Q \right )-e_{32} \left (\e_1\tcdot Q \right )-e_{21} \left (\e_3\tcdot Q \right )
 = 0\, .
 }
 For the terms of order $\alpha'$, first notice that by using momentum conservation and on-shell conditions, we have the identity:
 \ea{
 &\left (\e_1\tcdot K_2\right)\left (\e_3\tcdot K_1\right)\left (\e_2\tcdot Q \right )
+ 
\left (\e_3\tcdot Q \right )\left (
\left (\e_1\tcdot K_2\right)\left (\e_2\tcdot K_3\right)
- \left(\e_1\tcdot K_3\right) \left (\e_2\tcdot K_1\right) \right )
\nonumber \\[2mm]
&=
-\left (\e_1\tcdot K_3\right)\left (\e_3\tcdot K_1\right)\left (\e_2\tcdot Q \right )
-
\left (\e_3\tcdot Q \right )\left (
\left (\e_1\tcdot K_2\right)\left (\e_2\tcdot K_1\right)
- \left(\e_1\tcdot K_2\right) \left (\e_2\tcdot K_1\right) \right )
\nonumber \\
&\quad -\left (\e_1\tcdot Q\right)\left (\e_3\tcdot K_1\right)\left (\e_2\tcdot Q \right )
-
\left (\e_3\tcdot Q \right )\left (
\left (\e_1\tcdot K_2\right)\left (\e_2\tcdot Q\right)
- \left(\e_1\tcdot Q\right) \left (\e_2\tcdot K_1\right) \right ).
}
After the manipulation, the second term in the first line is zero identically. The terms in the last line are of order $q^2$. Consider the first term, which we now cyclic and reversal symmetrize:
\ea{
&-\left (\e_1\tcdot K_3\right)\left (\e_3\tcdot K_1\right)\left (\e_2\tcdot Q \right )
-\left (\e_2\tcdot K_1\right)\left (\e_1\tcdot K_2\right)\left (\e_3\tcdot Q \right )
-\left (\e_3\tcdot K_2\right)\left (\e_2\tcdot K_3\right)\left (\e_1\tcdot Q \right )
\nonumber \\
&
+\left (\e_1\tcdot K_3\right)\left (\e_3\tcdot K_1\right)\left (\e_2\tcdot Q \right )
+\left (\e_2\tcdot K_3\right)\left (\e_3\tcdot K_2\right)\left (\e_1\tcdot Q \right )
+\left (\e_1\tcdot K_2\right)\left (\e_2\tcdot K_1\right)\left (\e_3\tcdot Q \right )
\nonumber \\
&= 0 \, .
}
The terms add up to zero, and thus there are no terms linear in $q$  entering in the cyclic and reversal symmetrized expression. 
Finally, let us rewrite the terms of order $q^2$ so that they are manifestly cyclic and reversal symmetric.
It is easiest to first rewrite them once again using momentum and on-shell conditions,
\ea{
-\left (\e_1\tcdot Q\right)\left (\e_2\tcdot Q \right ) \left (\e_3\tcdot K_1\right)
-\left (\e_2\tcdot Q\right) \left (\e_3\tcdot Q \right )\left (\e_1\tcdot K_2\right)
- \left (\e_3\tcdot Q \right ) \left(\e_1\tcdot Q\right) \left (\e_2\tcdot K_3\right),
}
which is already cyclic symmetric. Manifesting also reversal symmetry finally gives:
\ea{
&
-\left (\e_1\tcdot Q\right)\left (\e_2\tcdot Q \right ) \e_3\tcdot K_{12}
-\left (\e_2\tcdot Q\right) \left (\e_3\tcdot Q \right )\e_1\tcdot K_{23}
- \left (\e_3\tcdot Q \right ) \left(\e_1\tcdot Q\right) \e_2\tcdot K_{31},
}
where $K_{ij} = \frac{K_i - K_j}{2}$.
 
 The final result of the integration, where cyclic and reversal symmetry have been preserved in fixing the M\"obius symmetry, thus reads
 \ea{
 \widetilde{M}_3^{\a'}= &\,
\kappa_d   \Bigg[
 \frac{A_3^{\a'} (1,2,3)-A_3^{\a'}(3,2,1)}{ \sqrt{\a'}}   
 \nonumber \\
&
- \frac{\a'}{\sqrt{2}} \, q_\mu q_\nu \left ( 
\e_1^\mu \e_2^\nu \e_3 \cdot k_{12} 
+\e_2^\mu \e_3^\nu \e_1 \cdot k_{23} 
+ \e_3^\mu \e_1^\nu \e_2 \cdot k_{31} 
\right ) \Bigg ]^2 \, ,
\nonumber \\[3mm]
=&  \frac{\kappa_d}{\sqrt{\a'}} \mathcal{C}_3 [A_3^{\a'} ]^2 + \Ord(\a' q^2 )
 }
 where $A_3^{\a'}$ is given in \Eq{A3} and $k_{ij} = \sfrac{k_i - k_j}{2}$.
 Notice that the leading term of order $q^0$ upon squaring the bracket is a manifestly permutation symmetric expression for $M_3^{\a'}$, i.e. $\mathcal{C}_3[A_3^{\a'}]^2$ (cf. below \Eq{eq:KM3}). Notice also that in the field theory limit $\a'\to 0$ or up to order $q^2$, one exactly gets $\mathcal{C}_3[A_3]^2$, respectively $\mathcal{C}_3[A_3^{\a'}]^2$, up to normalization.
  Thus, $\mathcal{C}_3[A_3^{\a'}]^2$ represents one consistent input in \Eq{softtheorem}
for calculating the soft limit of the dilaton in scattering processes with three other massless closed strings, 
as a consequence of ensuring cancellation of all large $z$ dependences. 
As discussed after \Eq{eq:KM3}, this expression consistently gives the correct, vanishing field theory limit of the soft dilaton scattering with three gravitons.
 
 %%%%%%%%%%%%%%%%%%%%%%

\section{On Gauge Invariance of the Dilaton Soft Theorem}
\label{App:gaugeinvariance}
In this appendix we discuss how gauge invariance is ensured in the expressions \Eq{softtheorem} and \Eq{DilatonSoftTheorem} at  order $q$. 
For this, we consider the commutator of the soft operators with the operator of gauge transformations $W_i = k_i \cdot \partial_{ \e_i }$.

The order-$q$ operators in \Eq{softtheorem} can be expressed as
\ea{
S_q^{(1)} = &- \frac{\varepsilon_{\mu \nu}}{2} \sum_{i=1}^n
\frac{  q_\rho q_\sigma}{k_i \cdot q} \left [
 J_i^{\mu \rho}  J_i^{\nu \sigma}
- \eta^{\mu \nu}  \left (
k_i^\rho
\partial_{k_{i}}^{{\sigma}} 
+
\e_i^\rho
\partial_{\e_{i}}^{{\sigma}} 
\right )
\right ] \, .
}
It can be checked that the two terms separately commute with $W_i$.
The second term is easy to check due to the linearity of the operators
\ea{
\left [W_i \, , \ k_i^\rho
\partial_{k_{i}}^{{\sigma}} 
+
\e_i^\rho
\partial_{\e_{i}}^{{\sigma}}  \right ]
= - k_i^\rho \partial_{\e_{i}}^{{\sigma}} +  k_i^\rho \partial_{\e_{i}}^{{\sigma}}  = 0 \, .
}
The first term is more involved. To understand the cancellations taking place, let us consider each component of the operator
\ea{
\varepsilon_{\mu \nu}
q_\rho J_i^{\mu \rho} q_\sigma J_i^{\nu \sigma}
= 
\varepsilon_{\mu \nu}q_\rho q_\sigma
\left ( L_i^{\mu \rho} L_i^{\nu \sigma} + S_i^{\mu \rho} S_i^{\nu \sigma} + 2 L_i^{\mu \rho} S_i^{\nu \sigma} \right ),
}
where in the last term we used the symmetric contractions in $\mu \nu$ and $\rho \sigma$ to sum two terms.
Here $L_i$ and $S_i$ are given by
\begin{equation}
L_i^{\mu \nu} = i \left( k_i^\mu \partial_{ k_{i}}^{ \nu} - k_i^\nu\partial_{ k_{i}}^{\mu} \right ) \, , \quad
S_i^{\mu \nu} = i \left( \e_i^\mu \partial_{ \e_{i}}^{ \nu} - \e_i^\nu\partial_{ \e_{i}}^{\mu} \right ).
\end{equation}
Using that $q^\mu \varepsilon_{\mu \nu}=0$, one then finds the following commutators:
\ea{
\left [W_i \, , \ 
\varepsilon_{\mu \nu}q_\rho q_\sigma
 L_i^{\mu \rho} L_i^{\nu \sigma} \right ]
 &= \varepsilon_{\mu \nu} q^\sigma \left ((k_i\cdot q) \eta^{\mu \nu} \partial_{ \e_i}^{\sigma}  - i 2 q_\rho L_i^{\mu \rho} k_i^{[\nu } \partial_{ \e_i}^{\sigma]} \right ),
 \\
 \left [W_i \, , \ 
\varepsilon_{\mu \nu}q_\rho q_\sigma
 S_i^{\mu \rho} S_i^{\nu \sigma} \right ]
 &= \varepsilon_{\mu \nu} q^\sigma \left (  (k_i\cdot q) \eta^{\mu \nu} \partial_{ \e_i}^{\sigma}+ i 2 q_\rho S_i^{\mu \rho} k_i^{[\nu } \partial_{ \e_i}^{\sigma]}  \right ),
 \\
 \left [W_i\, , \ 
2 \varepsilon_{\mu \nu}q_\rho q_\sigma
 L_i^{\mu \rho} S_i^{\nu \sigma} \right ]
 &= - 2 \varepsilon_{\mu \nu} q^\sigma \left ( (k_i\cdot q) \eta^{\mu \nu}\partial_{ \e_i}^{\sigma} 
 + i q_\rho\left ( L_i^{\mu \rho} k_i^{[\nu } -  S_i^{\mu \rho} k_i^{[\nu } \right )k_i^{[\nu } \partial_{ \e_i}^{\sigma]} \right ),
  }
where the notation $a^{[\mu}b^{\nu]} = a^\mu b^\nu - a^\nu b^\mu$ was used. The sum of the three commutators vanishes, such that
\ea{
\left [W_i \, , \ 
\varepsilon_{\mu \nu}q_\rho q_\sigma
 J_i^{\mu \rho} J_i^{\nu \sigma} \right ]
 &= 0 \, .
}
Notice that symmetry and transversality of $\varepsilon_{\mu \nu}$ were used to obtain these relations, but tracelessness was not necessary.

Now, consider the soft dilaton operator at order $q$:
\ea{
S_{\text{dilaton}}^{(1)}
=
 q_\mu \op{K}_0^\mu
+ \sum_{i=1}^{n} \frac{q_\rho q^\sigma}{ 2 k_i\cdot q} \left(      
 S_i^{ \rho \mu  }  S_{i, {\mu \sigma } }
  +   d     \e_{i}^{\rho} 
\partial_{\e_{i }}^{{\sigma}} 
-2\epsilon_i^\rho \partial_{k_i}^{\sigma}
\left (k_i \cdot \partial_{\e_{i }} \right )
\right) \, .
\label{S1dilaton}
}
The dilaton projector used to obtain this operator, is symmetric and transverse, thus it follows from the preceding discussion that the operator in this form also commutes with $W_i$.

To disentangle the different contributions to this operator, we need to understand how the different parts commute with $W_i$. 
First, notice that we can rewrite
\ea{
\frac{q_\rho q^\sigma}{ 2 k_i\cdot q} \left(      
 S_i^{ \rho \mu  }  S_{i, {\mu \sigma } }
  +   d     \e_{i}^{\rho} 
\partial_{\e_{i }}^{{\sigma}} \right )
= 
\frac{q^\rho q^\sigma}{  k_i\cdot q} \left(      
\e_i^\rho \partial_{\e_i}^\sigma \left [2 - \e_i \cdot \partial_{\e_i}\right ] +
\sfrac{1}{2}\e_i \cdot \e_i \partial_{\e_i}^\rho\partial_{\e_i}^\sigma + \sfrac{1}{2}\e_i^\rho \e_i^\sigma \partial_{\e_i}^2 \right ),
\label{threepointonshell}
}
where $q^2 = 0$ was used. Notice that the first term involving $\left [2 - \e_i \cdot \partial_{\e_i}\right ]$ vanishes when acting on graviton amplitudes, because of their quadratic dependence on $\e_i$. It is instructive to not yet use this property.
We consider the commutator of each term:
\ea{
\left [ W_i \, , \frac{q \cdot \e_i}{  k_i\cdot q}
q \cdot \partial_{\e_i} \left (2 - \e_i \cdot \partial_{\e_i}\right )
\right ] &= q \cdot \partial_{\e_i} \left (2 - \e_i \cdot \partial_{\e_i}\right ) - 
\frac{\e_i \cdot q  }{ k_i\cdot q} q\cdot \partial_{\e_i} W_i,
\\
\left [ W_i \, , 
\frac{\e_i \cdot \e_i}{2 k_i \cdot q} (q \cdot  \partial_{\e_i})^2 \right ]
&=
\frac{k_i \cdot \e_i}{ k_i \cdot q} (q \cdot  \partial_{\e_i})^2  = 0,
\\
\left [ W_i \, ,  \frac{(q \cdot \e_i)^2}{ 2 k_i\cdot q}\partial_{\e_i}^2 \right ]
&=\frac{q\cdot \e_i}{2} \partial_{\e_i}^2,
\\
\left [W_i\, , \ K_{0}^\mu \right ]
&= W_i \partial_{k_i}^{\mu} + \left (\e_i \cdot \partial_{\e_{i}} \right)\partial_{ \e_{i}}^{{\mu}}- \e_i^\mu \partial_{\e_{i}}^2,
\\
\left [W_i\, , \ - \frac{q \cdot \e_i}{ k_i\cdot q} (q \cdot  \partial_{ k_i} )W_i \right ]
&= q_\mu \left ( 
\frac{(\e_i \cdot q)  }{ k_i\cdot q} W_i \partial_{\e_i}^{\mu} 
- W_i \partial_{k_i}^{\mu} -  \partial_{\e_i}^{\mu } \right ) .
}
Apart from the second commutator, where $k_i \cdot \e_i = 0$ was used, the vanishing of the sum of terms is rather entangled, and are at this level not separable. However, we are considering these operators and commutators on graviton amplitudes, and thus it is only really relevant to understand the \emph{effective} commutators on the stripped amplitude $M_n = \e_i^\alpha \e_i^\beta M_{n, \alpha \beta}^{(i)}$.
Assuming the following properties for $M_{n, \alpha \beta}^{(i)}$:
\ea{
k_i^\alpha M_{n, \alpha \beta}^{(i)} = k_i^\beta M_{n, \alpha \beta}^{(i)} = \eta^{\alpha \beta} M_{n, \alpha \beta}^{(i)} = 0 \, ,
}
valid for tree-level graviton amplitudes, we find 
\ea{
\left [ W_i \, , \frac{q \cdot \e_i}{  k_i\cdot q}
q \cdot \partial_{\e_i} \left (2 - \e_i \cdot \partial_{\e_i}\right )
\right ] M_n &= 0,
\\
\left [ W_i \, ,  \frac{(q \cdot \e_i)^2}{ 2 k_i\cdot q}\partial_{\e_i}^2 \right ] M_n
&=0,
\\
\left [W_i\, , \ K_{0}^\mu \right ]
&= \partial_{k_i}^{\mu}W_i M_n,
\\
\left [W_i\, , \ - \frac{q \cdot \e_i}{ k_i\cdot q} (q \cdot  \partial_{ k_i} )W_i \right ]
&= - q_\mu\partial_{k_i}^{\mu}W_i M_n.
}
This explains why we can consider the terms in \Eq{threepointonshell} separately from the rest of the terms in \Eq{S1dilaton}. The last two commutators, however, do not separately vanish, unless $M_n$ is manifestly gauge invariant. Instead, the sum ensures commutation:
\ea{
\left [W_i\, , \ q_\mu K_{ 0}^\mu - \frac{q \cdot \e_i}{ k_i\cdot q} (q \cdot  \partial_{ k_i} )W_i \right ]M_n = 0 \, .
\label{gaugeinvariance}
}
This exposes a potential problem in separating the contribution of those two operators in the soft theorem, as done in the main text. The issue, however, also appears for Yang--Mills amplitudes in $d=4$, where the commutator has exactly the same deficiency, i.e.
\ea{
\left [W_i\, , \ K_{\Delta}^\mu \right ] A_n \stackrel{\atopfrac{ d=4}{\Delta = 1}}{=} \partial_{k_i}^{\mu}W_i A_n  \, .
}

We have argued in the main text that the operator which is non-local in $q$ in \Eq{gaugeinvariance} is not a real order $q$ contribution to the soft behavior of the dilaton as it can be resummed to zero.
Regarding the gauge invariance, this makes sense if we can replace that zero with another zero, which also ensures gauge invariance. This is, in fact, possible, i.e.
\ea{
\left [W_i\, , \ q_\mu K_{ 0}^\mu + (2- \e_i \cdot \partial_{\e_i}) q \cdot \partial_{k_i} \right ] M_n = 0 \, .
\label{gaugeinvariance2}
}
The new replacing term obviously annihilates the graviton amplitude, since
\ea{
(2- \e_i \cdot \partial_{\e_i}) M_n = 0 \, ,
}
for any $i=1, \ldots, n$. 
Since now the replacing term is local in $q$, we can even disregard it, i.e.
\ea{
\left [W_i\, , \ K_{ 0}^\mu + (2- \e_i \cdot \partial_{\e_i})\partial_{k_i}^\mu \right ] M_n = 0 \, .
}
An equivalent relation exists in the Yang--Mills case by replacing the factor $2$ with $1$ (in $d=4$ and for $\Delta=1$). 
We have thus argued that $K_0^\mu$ \emph{effectively} commutes with $W_i$ in this generalized sense. Thus it seems to be possible to disentangle the two contributions in \Eq{gaugeinvariance}, as done in the main text. It is plausible that the deficiency of $K_{\Delta=1}^\mu$ not immediately annihilating Yang--Mills amplitudes in $d=4$, or $K_0^\mu$ not immediately annihilating graviton amplitudes for any $d$,
 is related to the subtleties exposed here.

%%%%%%%%%%%%%%%%%%%%%%

\section{Conformal Symmetry of Two-Dilaton-Two-Graviton Amplitude}
\label{App:2dil2grav}

By replacing two polarization tensors with two dilaton projectors in the KLT expression for the four-graviton field theory amplitude, one gets the two-dilaton-two-graviton field theory amplitude. The corresponding $\delta$-stripped amplitude can be expressed as
\ea{
M_{2\phi 2g}=
\frac{16 \left(t k_1\cdot \epsilon _3 k_2\cdot \epsilon _4+u k_1\cdot
   \epsilon _4 k_2\cdot \epsilon _3+t u \epsilon _3\cdot \epsilon_4\right){}^2}{s t u}   \, ,
}
where labels 1 and 2 denote the dilatons and label 3 and 4 denote the gravitons.
The Mandelstam variables here read
\ea{
s = k_1 \cdot k_2 \, , \ t = k_2 \cdot k_3 \, , \ u =  - s - t \, .
}
Notice that the $\delta$-function has been stripped off according to \Eq{MomPoincare} by imposing the constraints $k_4 = - k_1 - k_2 - k_3$, $k_3 \cdot \e_4 = - (k_1+k_2)\cdot \e_4$, and $k_1 \cdot k_3 = - k_1 \cdot k_2 - k_2 \cdot k_3$.

This expression is \emph{manifestly} permutation symmetric in labels 1 and 2, but not in 3 and 4.
This lack of manifest permutation symmetry, shows up also when we consider the action
\ea{
\op{K}_0^\mu M_{2\phi 2g} =
\epsilon_3^\mu F_3 +  \epsilon_4^\mu F_4,
\label{KM2d2g}
}
with
\ea{
F_3 &=-32 \frac{ k_2\cdot \epsilon _4}{u} \left(\frac{ t k_1\cdot \epsilon _3
   k_2\cdot \epsilon _4+u k_1\cdot \epsilon _4 k_2\cdot \epsilon _3+t u
   \epsilon _3\cdot \epsilon _4}{s } \right)\, ,
 \\
 F_4&=  
 32 \frac{ k_2\cdot \epsilon
   _3 }{ t}
\left(   \frac{ t k_1\cdot \epsilon _3 k_2\cdot \epsilon _4+u k_1\cdot \epsilon _4
   k_2\cdot \epsilon _3+t u \epsilon _3\cdot \epsilon _4}{s} \right) \, ,
   }
i.e.\ the right-hand side of \Eq{KM2d2g} is not vanishing, nor is it permutation symmetric (not even on upon imposing 
momentum conservation and on-shell conditions). 
But if we enforce permutation symmetry on \Eq{KM2d2g} it is easy to see that 
\ea{
\op{K}_0^\mu \big [M_{2\phi 2g}(1,2,3,4) + M_{2\phi 2g}(1,2,4,3)\big ] =
\op{K}_0^\mu M_{2\phi 2g} + \left (\hat{K}_0^\mu M_{2\phi 2g}\big |_{3 \leftrightarrow 4}\right ) = 0 \, .
}
One can think of enforcing permutation symmetry simply as a means to restore the symmetry which is otherwise `lost' due to momentum-conservation not commuting with $\op{K}_0^\mu$.
Since $\op{K}_0^\mu$ does commute with permutation symmetrization, we could also have started with manifesting the $3 \leftrightarrow 4$ permutation symmetry in $M_{2\phi 2g}$. 

%%%%%%%%%%%%%%%%%%%%%%%%%%%%%%%%%%%%%%%%%%%%%%%%%%%%%%%%%%%%%%%%%%%%%%%%%%%
%%%%%%%%%%%%%%%%%%%%%%%%%%%%%%%%%%%%%%%%%%%%%%%%%%%%%%%%%%%%%%%%%%%%%%%%%%%
\bibliographystyle{nb}
\bibliography{SymGravAmps}

\end{document}